\PassOptionsToPackage{dvipsnames}{xcolor}
\documentclass[12pt]{article}

\usepackage[margin=2.5cm]{geometry}
\usepackage[utf8]{inputenc}
\DeclareUnicodeCharacter{03C0}{$\pi$}

\usepackage{amsthm,amsmath}
\usepackage{amssymb,amscd}
\usepackage{bbm}
\usepackage{amsfonts}
\usepackage{booktabs}
\usepackage{tabularx}
\usepackage{siunitx}
\usepackage{graphicx}
\usepackage{colortbl}
\usepackage{multirow}
\usepackage{array}
\usepackage{pdflscape}
\usepackage{afterpage}

\usepackage{lmodern}

\usepackage{hyperref}
\usepackage{url}

\usepackage[acronym,nomain,nohypertypes=all]{glossaries}
\glsdisablehyper

\newacronym{nlme}{NLME}{nonlinear mixed-effects}
\newacronym{ode}{ODE}{ordinary differential equation}
\newacronym{ude}{UDE}{universal differential equation}
\newacronym{hmm}{HMM}{hidden Markov model}
\newacronym{em}{EM}{expectation maximization}
\newacronym{saem}{SAEM}{stochastic approximation expectation maximization}
\newacronym{focei}{FOCEI}{first-order conditional estimation with interaction}
\newacronym{mcmc}{MCMC}{Markov chain Monte Carlo}
\newacronym{mcem}{MCEM}{Monte Carlo expectation maximization}
\newacronym{vpc}{VPC}{visual predictive check}
\newacronym{elbo}{ELBO}{evidence lower bound}
\newacronym{wres}{WRES}{weighted residuals}
\newacronym{pred}{PRED}{population predictions}
\newacronym{ipred}{IPRED}{individual predictions}
\newacronym{ebe}{EBE}{empirical Bayes estimate}
\newacronym{kde}{KDE}{kernel density estimate}
\newacronym{vbgf}{VBGF}{von Bertalanffy growth function}
\newacronym{lbfgs}{L-BFGS}{limited-memory Broyden-Fletcher-Goldfarb-Shanno}
\newacronym{lhs}{LHS}{Latin hypercube sampling}
\newacronym{pit}{PIT}{probability integral transform}
\newacronym{nuts}{NUTS}{no-U-turn sampler}
\newacronym{mle}{MLE}{maximum likelihood}
\newacronym{cv}{CV}{cross-validation}
\newacronym{pk}{PK}{pharmacokinetic}
\newacronym{pd}{PD}{pharmacodynamic}
\newacronym{inr}{INR}{international normalized ratio}
\newacronym{rmse}{RMSE}{root mean squared error}

\usepackage{natbib}
\usepackage{tikz}
\usetikzlibrary{positioning,arrows.meta}

\usepackage{fancyvrb}
\DefineVerbatimEnvironment{Code}{Verbatim}{}
\DefineVerbatimEnvironment{CodeInput}{Verbatim}{fontshape=sl}
\DefineVerbatimEnvironment{CodeOutput}{Verbatim}{}

\let\proglang=\textsf
\newcommand{\pkg}[1]{{\fontseries{b}\selectfont #1}}
\makeatletter
\newcommand\code{\bgroup\@makeother\_\@makeother\~\@makeother\$\@codex}
\def\@codex#1{{\normalfont\ttfamily\hyphenchar\font=-1 #1}\egroup}
\makeatother

\DeclareMathOperator*{\argmax}{arg\,max}

\newcommand{\nl}{\pkg{NoLimits.jl}}
\newcommand{\Ex}{\mathbb{E}}

\newcolumntype{L}[1]{>{\raggedright\let\newline\\\arraybackslash\hspace{0pt}}m{#1}}
\newcolumntype{C}[1]{>{\centering\let\newline\\\arraybackslash\hspace{0pt}}m{#1}}
\newcolumntype{R}[1]{>{\raggedleft\let\newline\\\arraybackslash\hspace{0pt}}m{#1}}

\usepackage{lineno}

\title{\textbf{\pkg{NoLimits.jl}: Flexible and Composable Nonlinear
Mixed-Effects Modeling in Julia}}
\author{Manuel Huth\,$^{\text{1,2}}$,
Jonas Arruda\,$^{\text{1,2}}$,
Nina Schmid\,$^{\text{1,2}}$,
Roy Gusinow\,$^{\text{1,2}}$,\\
Vincent Wieland\,$^{\text{1,2}}$,
Clemens Peiter\,$^{\text{1,2}}$, and
Jan Hasenauer\,$^{\text{1,2,}\ast}$}
\date{}

\begin{document}

\maketitle
{\small
$^{\text{1}}$ Life and Medical Sciences (LIMES) Institute, University of Bonn, Bonn, Germany\\
$^{\text{2}}$ Bonn Center for Mathematical Life Sciences, University of Bonn, Bonn, Germany\\
$^\ast$ Corresponding author (jan.hasenauer@uni-bonn.de)
}

\section*{Abstract}
Nonlinear mixed-effects models are widely used to analyze longitudinal data, but existing open-source software often supports only a limited subset of the model structures, inference methods, machine-learning components, automatic differentiation techniques, and random-effects distributions required in modern applications. We introduce \pkg{NoLimits.jl}, an open-source \proglang{Julia} package for flexible and composable nonlinear mixed-effects modeling. Its macro-based modeling language enables observation and latent-state models to be constructed from diverse building blocks, including ordinary differential equations, Markov models, and neural networks. \pkg{NoLimits.jl} supports flexible, covariate-dependent observation and random-effects distributions and provides a unified interface to frequentist inference through Laplace approximation, stochastic expectation maximization, and Bayesian Markov chain Monte Carlo methods. We demonstrate the package on three case studies showcasing its workflows, integration of differentiable machine-learning components, and data-driven estimation of random-effects distributions using normalizing flows. Together, these capabilities substantially expand the range of nonlinear mixed-effects models that can be specified, estimated, and compared within a single open-source framework.

\noindent\textbf{Keywords:} nonlinear mixed-effects models, longitudinal data analysis, scientific machine learning, normalizing flows, \proglang{Julia}.

\section{Introduction}

\Gls{nlme} models are a central tool across a wide range of scientific domains, including pharmacometrics \citep{mould2012basic, gobburu2010pharmacometrics}, clinical data \citep{maheux2023forecasting, cogswell2023evidence}, systems biology\citep{karlsson1995three, frohlich2017scalable}, psychometrics\citep{rijmen2003nonlinear, chung2021cross}, and ecology \citep{schliehe2012application, muff2020accounting}. Such studies generate repeated measurements from multiple individuals, whose responses often vary substantially due to both observed and unobserved sources of heterogeneity \citep{lavielle2014mixed,pinheiro2007linear}. Throughout this article, we use the term \emph{individual} generically to denote any experimental unit with repeated observations, including humans, animals, cells, countries, or other subjects under study.
\Glspl{nlme} address the variability by combining a population-level model with individual-specific random effects, enabling the joint estimation of population characteristics and individual deviations. Their hierarchical structure further supports the integration of sophisticated model components, ranging from mechanistic systems such as differential equations \citep{mould2012basic} to machine-learning-based approaches \citep{ngufor2019mixed,sigrist2022latent}. 
Moreover, we view hierarchical Bayesian models as \gls{nlme} models, with the Bayesian formulation differing primarily by the use of prior distributions and posterior inference \citep{gelman2013bayesian}.

At the same time, the complexity of models used in practice has grown substantially. Modern applications increasingly integrate mechanistic models with neural networks \citep{rackauckas2020accelerated}, particularly in fields such as neuroscience \citep{el2025universal} and systems biology \citep{philipps2025current}, as well as with other differentiable machine-learning components \citep{wortwein2023neural}. While these developments expand the range of problems that can be represented, they also introduce new computational challenges which need to be addressed by scientific software. Researchers require frameworks that can integrate different model components, probability distributions, and inference algorithms within a unified workflow, while remaining compatible with modern automatic differentiation and machine-learning ecosystems.

Existing software systems differ substantially in their modeling abstractions, inference capabilities, extensibility, and open-source availability. General mixed-effects frameworks such as \pkg{nlme} \citep{pinheiro2000mixed} and \pkg{lme4} \citep{bates2015lme4} primarily target reduced-form mixed-effects formulations and do not natively support mechanistic \gls{ode} systems or Markov models and are focused on Gaussian random effects. 
Dedicated \gls{nlme} software, including proprietary platforms such as \pkg{NONMEM} \citep{bauer2019nonmem}, \pkg{Monolix} \citep{Monolix2024R1}, and \pkg{Pumas} \citep{rackauckas2020accelerated}, as well as open-source alternatives such as \pkg{nlmixr2} \citep{fidler2025nlmixr2} and \pkg{saemix} \citep{comets2017parameter}, provides specialized workflows for \acrshort{ode}-based mechanistic modeling. However, support for Markov-model and latent-state formulations remains comparatively limited in these frameworks. Specialized tools such as \pkg{mHMMbayes} \citep{mHMMbayes} instead focus specifically on hierarchical \glspl{hmm}. In parallel, probabilistic programming frameworks such as \pkg{Stan} \citep{stan}, \pkg{Turing.jl} \citep{turing,turingjl}, and \pkg{PyMC} \citep{pymc2023} offer flexible Bayesian inference workflows and allow users to define custom hierarchical \gls{ode} and latent-state models, but they often require substantial manual implementation of \acrshort{nlme}-specific workflows.

These frameworks also differ in their support of standard inference procedures. \pkg{NONMEM} and \pkg{nlmixr2} support estimation by the \gls{saem}~\citep{wei1990monte, kuhn2004coupling} and \gls{focei}~\citep{pinheiro2006efficient,wang2007derivation}, whereas \pkg{Monolix} and \pkg{saemix} are centered on \gls{saem} estimation. 
The probabilistic programming languages \pkg{Stan}, \pkg{Turing.jl}, and \pkg{PyMC} emphasize Bayesian \gls{mcmc} and variational-inference approaches \citep{blei2017variational}. Among dedicated \gls{nlme} frameworks, \pkg{Pumas} provides one of the broadest collections of inference paradigms, including \acrshort{focei}-, Laplace-, and \acrshort{saem}-based methods, as well as variational and Bayesian workflows.

Beyond model classes and inference algorithms, frameworks vary substantially in their support for random effects distributions  and integration with machine-learning libraries. In many classical \gls{nlme} frameworks, including \pkg{Monolix}, \pkg{saemix}, and \pkg{nlmixr2}, support for heavy-tailed or user-defined random-effects distributions remains limited, while \pkg{Pumas} offers greater flexibility. \pkg{Pumas}, \pkg{Turing.jl}, and \pkg{PyMC} provide more composable interfaces to modern neural-network ecosystems and automatic differentiation, whereas neural-network-based models in frameworks such as \pkg{NONMEM} and \pkg{Monolix} often require manual specification of network architectures \citep{bram2025iiv}. 

These limitations highlight the need for open-source \gls{nlme} frameworks that support diverse model classes and inference procedures while remaining compatible with modern automatic differentiation and machine-learning ecosystems. To the best of our knowledge, no existing open-source framework provides this combination of capabilities, and in particular none enables the integration of mixed \glspl{hmm} with a diverse range of built-in neural network components.

\begin{figure}[t!]
  \centering
  \includegraphics[width = \textwidth]{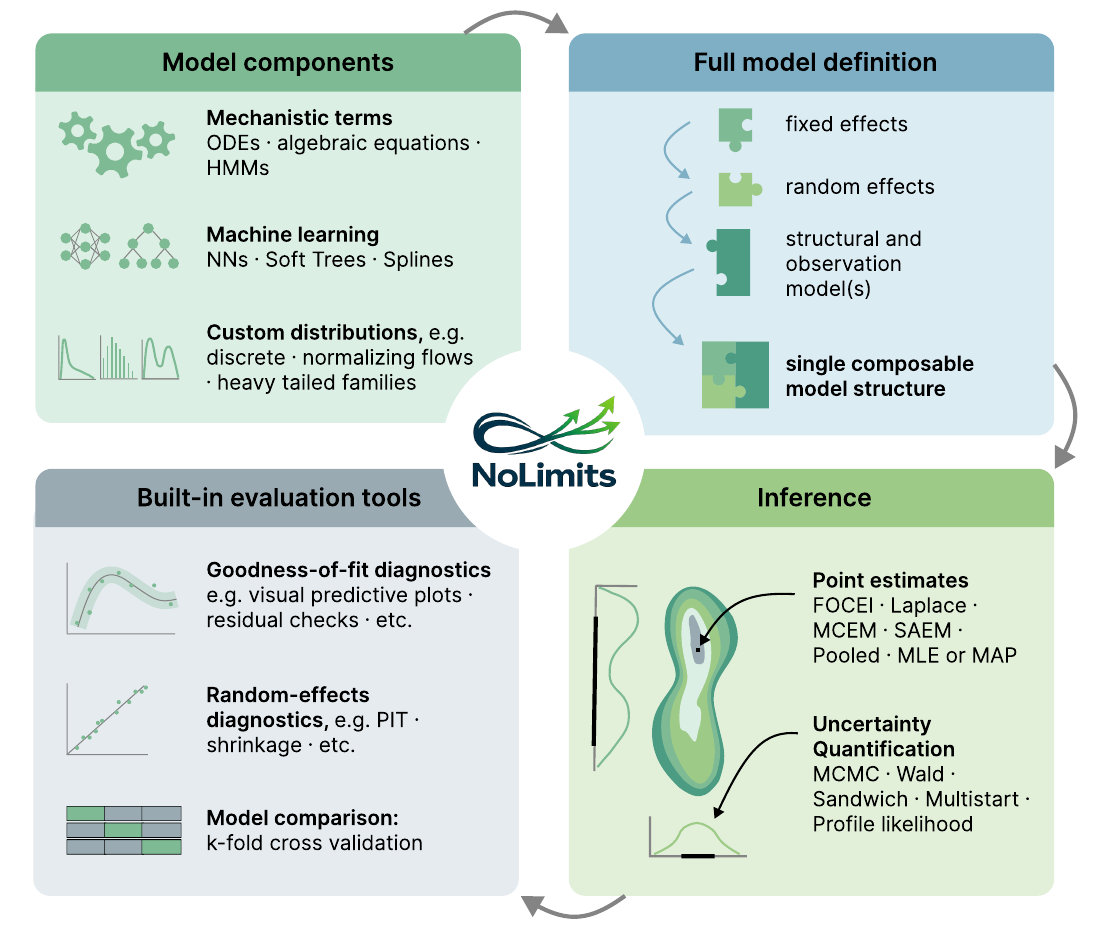}
  \caption{Overview of the \nl{} workflow. Mechanistic terms, machine-learning components, and custom distributions are combined into a single composable model structure comprising fixed effects, random effects, and structural and observation models. Inference yields estimates for fixed and random effects, optionally including uncertainty quantification. Built-in evaluation tools support model validation, comparison, and diagnostic assessment, guiding potential revisions to the selected model components.}
  \label{fig:overview}
\end{figure}

To address these limitations, we introduce \pkg{NoLimits.jl} (\textbf{No}n \textbf{Li}near \textbf{Mi}xed Effec\textbf{ts}.jl), an open-source framework for flexible nonlinear mixed-effects modeling in \proglang{Julia} \citep{bezanson2017julia}. \pkg{NoLimits.jl} unifies mechanistic modeling, statistical inference, and scientific machine learning within a common modeling and inference framework. 
In \nl{}, mechanistic terms,
machine-learning components, and custom distributions can be
composed into a single model definition combining fixed effects, random
effects, and structural and observation models. Point-estimation and uncertainty-quantification can be conducted with various methods and the fitted model assessed with goodness-of-fit, random-effect-specific, and
model-comparison tools (Figure~\ref{fig:overview}).

The core modeling abstraction in \nl{} represents observation models as parameterized conditional distributions and that inference is decoupled from model specification. The capabilities of \nl{} can be broadly divided into three areas: flexible model specification, integration with the \proglang{Julia} scientific machine-learning ecosystem, and support for a wide range of inference procedures.

Flexible model specification in \nl{} naturally encompasses a broad range of longitudinal models, including mechanistic models such as \gls{ode} systems \citep{hartman2002ordinary} and discrete- and continuous-time Markov models with hidden or observed states \citep{rabiner1989tutorial,zucchini2016hidden}, differentiable machine-learning models such as neural networks \citep{goodfellow2016deep,hornik1989multilayer} and soft trees \citep{irsoy2012soft}, hybrid mechanistic--neural models \citep{rackauckas2020universal}, and joint models with multiple observation types. To support these diverse model classes, \nl{} provides a macro-based language for composing observation models and latent structures from rich model components and probability distributions. This includes the rich collection of distributions available through \pkg{Distributions.jl} \citep{besancon2021distributions}, including heavy-tailed distributions.
The random-effects distribution can likewise be chosen flexibly from the
\pkg{Distributions.jl} ecosystem, also including univariate and multivariate,
heavy-tailed, and normalizing flows \citep{rezende2015variational} for flexible distributional modeling, provided through
\pkg{NormalizingFlows.jl} \citep{turing, turingjl}. In addition, the parameters of
the random-effects distribution may depend on covariates through arbitrary
differentiable functions, allowing the specification of both known and
previously unknown covariate--random-effects relationships.

The second capability area is tight integration with the \proglang{Julia} scientific computing ecosystem, enabling \nl{} to use established tools for machine learning, differential equations, optimization, and automatic differentiation. Neural-network-based model components can be implemented directly using \pkg{Lux.jl} \citep{pal2023lux} or \pkg{SimpleChains.jl} \citep{elrod2022simplechains}, enabling neural structural models, observation mappings, and random-effects distributions. Differentiable soft decision trees are additionally available as interpretable function approximators for each of these components. Mechanistic models leverage the SciML ecosystem through \pkg{OrdinaryDiffEq.jl} \citep{rackauckas2017differentialequations}, providing access to differentiable \gls{ode} solvers, event handling, and callback mechanisms. Automatic differentiation is supported through \pkg{ForwardDiff.jl} \citep{revels2016forward}, while numerical optimization is provided by \pkg{Optimization.jl} \citep{dixit2023optimization}, including optimizers from \pkg{Optimisers.jl} \citep{innes:2018} that are widely used in machine-learning applications.

The third capability area is support for a broad range of inference procedures through a common interface. These include Laplace-based approximation methods, stochastic \gls{em} algorithms such as \gls{saem} and \gls{mcem}, and Bayesian inference via \gls{mcmc} methods. Additionally, users can specify fixed-effects-only models and estimate parameters for them via \gls{mle}, \gls{mcmc}, or variational inference methods through \pkg{Turing.jl}. 
Because inference is decoupled from model specification, users can switch between estimation procedures with minimal changes to the analysis workflow, facilitating direct comparisons of alternative inference strategies under a common model.

\nl{} is released as open-source software under the MIT license and is developed publicly at \url{github.com/manuhuth/NoLimits.jl}, with documentation available at \url{manuhuth.github.io/NoLimits.jl}. The latest stable version can be installed from the \proglang{Julia} repository using the package manager.
\begin{CodeInput}
  using Pkg
  Pkg.add("NoLimits")
\end{CodeInput}
The package requires \proglang{Julia}~1.12 or later and can be installed using \code{Pkg.add}. 

The remainder of this article is organized as follows. Section~\ref{sec:nlme} reviews the \gls{nlme} modeling framework together with the estimation, uncertainty quantification, prediction, and evaluation methods underlying \pkg{NoLimits.jl}. Sections~\ref{sec:model_macro}, \ref{sec:est_nl}, and \ref{sec:uq_nl} describe model specification using the \code{@Model} macro, parameter estimation through a unified inference interface, and uncertainty quantification and model diagnostics, respectively. Section~\ref{sec:examples} demonstrates the framework on three case studies: an \gls{ode} model of warfarin pharmacokinetics, a learnable concentration-effect function for warfarin pharmacodynamics, and a normalizing flow as a flexible random-effects distribution for fish growth. Section~\ref{sec:comparison} compares \pkg{NoLimits.jl} with existing software frameworks, and Section~\ref{sec:conclusions} concludes.

\section{Nonlinear mixed-effects models}\label{sec:nlme}

In the following, we review the \gls{nlme} model class on which
\pkg{NoLimits.jl} is built. We present the
frequentist formulation first and add the Bayesian perspective as an extension in
Section~\ref{sec:bayes}. For clarity, we present the framework using a single grouping level, with one
random-effect vector per individual and a single, uncensored observation.
The formulation extends naturally to multiple nested and
crossed grouping levels (Appendix~\ref{app:nested}), multiple observation types per
individual (Appendix~\ref{app:multi}), and censored observations
(Appendix~\ref{app:censor}).

\subsection{The frequentist's mixed-effects model}

We begin with the frequentist formulation, in which the population parameters are fixed unknowns shared across all individuals. Individuals are indexed by $i = 1, \ldots, N$, where $N$ is the total number of individuals, and observations within individual $i$ are indexed by $j = 1, \ldots, n_i$, where $n_i$ is the number of observations for individual $i$. Both the number and timing of observations may vary across individuals.

\subsubsection{Hierarchical generative model.}
Let $\theta \in \Theta \subseteq \mathbb{R}^{n_\theta}$ denote the population
parameters shared across all individuals, and let
$x_i \in \mathbb{R}^{n_{x_i}}$ denote the time-invariant covariates of
individual $i$. The data-generating process is assumed to consist of two
levels \citep{lindstrom1990nonlinear, pinheiro2007linear},
\begin{align}
  B_i &\sim p_B(\cdot \mid x_i, \theta), \label{eq:nlme_re}\\
  Y_i \mid B_i = b &\sim p_{Y \mid B}(\cdot \mid b, x_i, \theta),
  \label{eq:nlme_obs}
\end{align}
where $B_i$ is the random-effect vector of individual $i$, taking values in
the support $\mathcal{B} \subseteq \mathbb{R}^{n_b}$ with density
$p_B(b \mid x_i, \theta)$, and $Y_i$ is the response vector of individual
$i$, taking values in the support
$\mathcal{Y}_i \subseteq \mathbb{R}^{n_i}$. Conditional on $B_i$, the
response is distributed according to the observation density
$p_{Y \mid B}(y \mid b, x_i, \theta)$. We write
$b_i \in \mathcal{B}$ and $y_i \in \mathcal{Y}_i$ for realizations of
$B_i$ and $Y_i$, respectively. Observations are, conditionally on covariates, assumed independent across individuals,
whereas observations within an individual are correlated through the shared
random effect $B_i$ and the conditional joint distribution $p_{Y \mid B}(\cdot \mid b, x_i, \theta)$.
As outlined in the notation, the random-effects distribution may depend on individual-fixed covariates $x_i$, allowing
individual heterogeneity modeled by random effects to vary according to baseline characteristics.

\subsubsection{Flexibility of the conditional model.}
The hierarchical formulation in Equations~\ref{eq:nlme_re} and \ref{eq:nlme_obs} restricts the
conditional model $p_{Y \mid B}(y \mid b, x, \theta)$ only by requiring its density to be evaluable,
so observation distributions may be continuous, binary, count, or censored, with multiple observations of
different types modeled jointly for the same individual. The structural model itself may be a
closed-form expression, the solution of a system of \glspl{ode}, the initial, transition or emission
structure of a Markov model with hidden or observed states, or a differentiable machine-learning
model such as a neural network or soft tree, and
subject to mild technical conditions these components combine freely within a single model.
\nl{} implements all of these structural forms through a common model syntax
(Section~\ref{sec:model_macro}).

\subsubsection{Flexibility of the random-effect distribution.}
The random-effects distribution is specified independently of the observation model. In addition to the multivariate Gaussian and log-normal families commonly used in \gls{nlme} modeling, the formulation accommodates arbitrary continuous univariate or multivariate distributions, including heavy-tailed and skewed families such as the Student's $t$, beta, $\chi^2$, and exponential distributions, as well as user-defined distributions and normalizing flows \citep{rezende2015variational}, which construct complex probability distributions by transforming a simple base distribution through a sequence of invertible differentiable mappings whose parameters are learned from data \citep{rezende2015variational,papamakarios2021normalizing}, enabling flexible random-effects distributions to be learned from data.
In \nl{}, we draw standard parametric families from \pkg{Distributions.jl} and
use learnable normalizing flows for distributions estimated from data
(Section~\ref{sec:model_macro}).

For normalizing flows, the random effect is generated as $B_i = T_{\psi}(U_i)$, where $U_i \sim p_U$ is a base random variable, typically a multivariate standard Gaussian, and $T_{\psi}$ is an invertible differentiable transformation with parameters $\psi$. The corresponding density follows from the change-of-variables formula,
\begin{equation}
p_B(b \mid \psi)
= p_U\bigl(T_{\psi}^{-1}(b)\bigr)
\left| \det \nabla_b T_{\psi}^{-1}(b) \right|,
\label{eq:flow}
\end{equation}
yielding a tractable likelihood in which $\psi$ can be estimated jointly with the remaining model parameters \citep{rezende2015variational,papamakarios2021normalizing}. Section~\ref{sec:flow} illustrates this approach using planar flows, which compose multiple such transformations \citep{rezende2015variational}.

\subsection{Joint and marginal likelihood}

The hierarchical structure in Equations~\ref{eq:nlme_re} and
\ref{eq:nlme_obs} induces the factorization
\begin{align}
  p_{Y,B}(y_i, b_i \mid x_i, \theta)
  =
  p_{Y \mid B}(y_i \mid b_i, x_i, \theta)\,
  p_B(b_i \mid x_i, \theta),
  \label{eq:joint}
\end{align}
which defines the joint density of the response and random effects for
individual $i$. When viewed as a function of $\theta$, this quantity is
commonly referred to as the \emph{complete-data likelihood}. If the random
effects were observed, inference would be based directly on the
corresponding complete-data log-likelihood,
\begin{equation}
  \ell_c(\theta)
  =
  \sum_{i=1}^{N}
  \log p_{Y,B}(y_i, b_i \mid x_i, \theta),
  \label{eq:cdll}
\end{equation}
which would be straightforward to evaluate.

In practice, however, the random effects are latent and therefore
unobserved. Inference must instead be based on the marginal distribution
of the observed responses, obtained by integrating the random effects out
of the joint density \citep{dempster1977maximum,lindstrom1990nonlinear},
\begin{equation}
  p_Y(y_i \mid x_i, \theta)
  =
  \int_{\mathcal{B}}
  p_{Y \mid B}(y_i \mid b, x_i, \theta)\,
  p_B(b \mid x_i, \theta)\,
  \mathrm{d}b,
  \label{eq:marginal}
\end{equation}
where we assume that the integral exists for every
$\theta \in \Theta$ \citep{pinheiro1995approximations}. Under
independence across individuals, the population parameters are estimated
by maximizing the marginal log-likelihood
\begin{equation}
  \hat{\theta}
  =
  \argmax_{\theta \in \Theta} \ell(\theta),
  \qquad
  \ell(\theta)
  =
  \sum_{i=1}^{N}
  \log p_Y(y_i \mid x_i, \theta),
  \label{eq:mll}
\end{equation}
or, equivalently, by minimizing the negative marginal log-likelihood.

For many practically relevant models, the integral in
Equation~\ref{eq:marginal} is not available in closed form. This occurs,
for example, when the conditional model is nonlinear in the random
effects, or non-Gaussian observation distributions are used
\citep{pinheiro1995approximations,lindstrom1990nonlinear}. Consequently,
the marginal log-likelihood in Equation~\ref{eq:mll} is often intractable,
and estimation methods must approximate the marginal integral. The
approaches described in Section~\ref{sec:point_est} differ primarily in
how this approximation is performed, through repeated evaluations,
approximations, or expectations involving the complete-data likelihood in
Equation~\ref{eq:joint}.
In \nl{}, a model specification determines this joint density directly, and the
estimators of Section~\ref{sec:est_nl} form the approximations to the otherwise
intractable marginal likelihood.

\subsection{Empirical Bayes estimates}\label{sec:ebe}

Although inference for the population parameters is based on the marginal
likelihood in Equation~\ref{eq:mll}, many applications additionally require
individual-specific estimates of the latent random effects. These estimates
are obtained from the conditional distribution of the random effects given
the observed response of an individual. By Bayes' theorem applied to
Equations~\ref{eq:nlme_re} and \ref{eq:nlme_obs}, this distribution is
given by
\begin{equation}
  p_{B \mid Y}(b \mid y_i, x_i, \theta)
  \propto
  p_{Y \mid B}(y_i \mid b, x_i, \theta)\,
  p_B(b \mid x_i, \theta),
  \label{eq:re_post}
\end{equation}
where the normalizing constant is the marginal density of
Equation~\ref{eq:marginal}. The \gls{ebe}
\citep{lindstrom1990nonlinear, davidian2003nonlinear} of
the random effects for individual $i$ is defined as the posterior mode,
\begin{equation}
  \hat{b}_i(\theta)
  =
  \argmax_{b \in \mathcal{B}}
  \left[
    \log p_{Y \mid B}(y_i \mid b, x_i, \theta)
    +
    \log p_B(b \mid x_i, \theta)
  \right].
  \label{eq:ebe}
\end{equation}

The \gls{ebe} plays a central role throughout mixed-effects modeling. It
serves as the expansion point for the Laplace and \gls{focei}
approximations discussed in Section~\ref{sec:point_est} and provides the
individual-level predictions and residual diagnostics.

\subsection{Point estimation}\label{sec:point_est}

To address the challenge posed by the intractable integral in
Equation~\ref{eq:marginal}, we review a spectrum of
well-established estimation methods that trade computational cost against
statistical accuracy: a pooled estimator that fixes the random effects and thus avoids integration over
random effects, the deterministic Laplace and \gls{focei}
approximations, and the stochastic \gls{mcem} and \gls{saem} algorithms
\citep{tierney1986accurate, wang2007derivation, wei1990monte,
delyon1999convergence}. The following sections describe these methods,
while the corresponding derivations are deferred to
Appendix~\ref{app:laplace}, Appendix~\ref{app:em}, and
Appendix~\ref{app:saem}.

In \nl{}, these estimators share a common fitting interface, so a single model
specification can be moved between them, or chained from an inexpensive pooled
fit to a Laplace, \gls{focei}, or \gls{saem} refinement, without being rewritten
(Section~\ref{sec:est_nl}).

\subsubsection{Pooled estimation.}
As a computationally inexpensive baseline, one may ignore
between-individual variability and estimate the population parameters
using a pooled model. The classical naive pooled estimator treats all
observations as arising from a single population and ignores
inter-individual variability \citep{sheiner1980evaluation}.
The classical naive pooled estimator treats all observations as arising
from a single population and ignores inter-individual variability
\citep{sheiner1980evaluation} by replacing each random effect with its conditional mean
$\mu_{b_i} = \Ex(B_i \mid x_i, \theta)$ and maximizing the resulting
population-level likelihood,
\begin{equation}
  \hat{\theta}
  =
  \argmax_{\theta \in \Theta}
  \sum_{i=1}^{N}
  \log p_{Y \mid B}
  \bigl(
    y_i \mid B = \mu_{b_i},
    x_i,
    \theta
  \bigr).
  \label{eq:pooled}
\end{equation}
For random-effects distributions without covariate effects,
Equation~\ref{eq:pooled} reduces to the classical naive pooled estimator.

Because no integration over random effects is required, pooled
estimation is computationally inexpensive and can serve as an
initializer for more sophisticated estimators or as a preliminary model
check. Since between-individual variability is ignored, however, the
parameters of the random-effects distribution
$p_B(b \mid x_i, \theta)$ are generally not identifiable.

\subsubsection{Laplace approximation.}
The Laplace approximation replaces the marginal integral in
Equation~\ref{eq:marginal} by a second-order Taylor expansion of the log
joint density around the \gls{ebe}
\citep{tierney1986accurate}. Because the gradient of the log joint density
vanishes at the mode, the resulting quadratic approximation yields a
Gaussian integral that can be evaluated analytically,
\begin{equation}
  p_Y(y_i \mid x_i, \theta)
  \approx
  (2\pi)^{n_b/2}
  \det\bigl(-H_i(\theta)\bigr)^{-1/2}
  p_{Y \mid B}(y_i \mid \hat{b}_i, x_i, \theta)
  p_B(\hat{b}_i \mid x_i, \theta),
  \label{eq:laplace}
\end{equation}
where $\hat{b}_i = \hat{b}_i(\theta)$ is the \gls{ebe} from
Equation~\ref{eq:ebe} and $H_i(\theta)$ is the Hessian of the log joint
density with respect to $b$ at $\hat{b}_i$. Each evaluation therefore requires
both the \gls{ebe} and its Hessian, and the approximation is accurate when
the posterior of Equation~\ref{eq:re_post} is close to Gaussian and the
random-effects dimension $n_b$ is moderate. A derivation, including bounded
random-effect domains and the gradient of the approximate marginal
log-likelihood via the envelope theorem, is given in
Appendix~\ref{app:laplace}.

\subsubsection{First-order conditional estimation with interaction.}
\Gls{focei} uses the same approximation as Equation~\ref{eq:laplace} and
differs only in its treatment of the curvature matrix $H_i(\theta)$.
Instead of the exact Hessian, it employs a first-order linearization of the
conditional model around the \gls{ebe}, approximating the curvature of the
conditional log-likelihood by its Fisher information \citep{fisher1925theory}
while retaining the exact curvature of the random-effects distribution
\citep{wang2007derivation,pinheiro2006efficient}. As a result, it
is often faster than the Laplace approximation and has become a standard
method in pharmacometrics, although its accuracy may degrade when the
conditional model is highly nonlinear in the random effects. The explicit
curvature matrix and a derivation are provided in Appendix~\ref{app:laplace}.

\subsubsection{Monte Carlo expectation maximization.}
The \gls{em} algorithm avoids direct optimization of the marginal
likelihood, alternating between an expectation and a maximization step
\citep{dempster1977maximum}. At iteration $t$ with current estimate
$\theta_t$, the expectation step constructs the surrogate objective, $Q$, which is the
posterior expectation of the complete-data log-likelihood of
Equation~\ref{eq:cdll} derived from the \gls{elbo},
\begin{equation}
  Q(\theta \mid \theta_t)
  =
  \sum_{i=1}^{N}
  \mathbb{E}_{B \sim
  p_{B \mid Y}(\cdot \mid y_i, x_i, \theta_t)}
  \left[
    \log p_{Y,B}(y_i, B \mid x_i, \theta)
  \right],
  \label{eq:qfun}
\end{equation}
and the maximization step updates the parameters according to
\begin{equation}
  \theta_{t+1}
  =
  \argmax_{\theta \in \Theta}
  Q(\theta \mid \theta_t).
\end{equation}

For nonlinear or non-Gaussian models, the expectation in
Equation~\ref{eq:qfun} is generally intractable because the posterior
distribution in Equation~\ref{eq:re_post} has no closed form.
\Gls{mcem} addresses this by replacing the expectation with a Monte Carlo
average over samples drawn from the posterior using \gls{mcmc}
\citep{wei1990monte}. Common choices include Metropolis--Hastings\citep{robert2004monte} and the
\gls{nuts}~\citep{hoffman2014no},
with Metropolis--Hastings often serving as a natural default because it
requires only evaluations of the joint density and no gradients of the
conditional model. As the number of Monte Carlo samples increases across
iterations, the limit points of the algorithm coincide with stationary
points of the marginal log-likelihood. Unlike the exact \gls{em}
algorithm, however, the stochastic expectation step does not guarantee
monotonic improvement of the objective \citep{wei1990monte}. Additional
details on the sampling identities are provided in
Appendix~\ref{app:em}.

\subsubsection{Stochastic approximation expectation maximization.}
For models in which each likelihood evaluation is expensive, for example when the conditional density
requires solving an \gls{ode}, drawing a new Monte Carlo sample at every iteration is costly. \gls{saem}
instead maintains a running stochastic approximation of the surrogate
\citep{delyon1999convergence, kuhn2004coupling}. At iteration $t$ it draws $M_t$ samples
$b_i^{(t,1)}, \ldots, b_i^{(t,M_t)}$ from the posterior of Equation~\ref{eq:re_post} and updates
\begin{equation}
  Q^{(t)}(\theta)
  = (1 - \gamma_t)\, Q^{(t-1)}(\theta)
    + \gamma_t \sum_{i=1}^{N} \frac{1}{M_t} \sum_{s=1}^{M_t}
      \log p_{Y, B}\bigl(y_i, b_i^{(t,s)} \mid x_i, \theta\bigr),
  \label{eq:saem}
\end{equation}
where the step sizes $\{\gamma_t\}$ satisfy the Robbins--Monro conditions
$\sum_t \gamma_t = \infty$ and $\sum_t \gamma_t^2 < \infty$. These
conditions ensure, under suitable convergence conditions, convergence of the SAEM iterates \citep{delyon1999convergence}. The schedule of
Appendix~\ref{app:gamma} is used in practice.
When the complete-data model belongs to an exponential family, \gls{saem}
reduces to recursive updates of finite-dimensional sufficient statistics,
followed by closed-form maximization steps
\citep{kuhn2005maximum,comets2017parameter}. Parameters that enter as
natural parameters of the exponential family can be updated analytically,
whereas parameters appearing in nonlinear or neural components are
optimized numerically. A single model fit can therefore combine closed-form
and numerical updates. The sufficient-statistic formulation is derived in
Appendix~\ref{app:saem}.

\subsection{Uncertainty quantification}\label{sec:uq}

While point estimation provides a single best-fitting value for each parameter, these estimates are associated with uncertainty because they are inferred from a finite and potentially noisy sample of data. In \nl{}, we implemented three
interval estimates for a likelihood-based fit, all evaluated at the
maximum-likelihood estimate of Equation~\ref{eq:mll}: Wald-type intervals
based on the observed information matrix, a misspecification-robust variant
based on the sandwich estimator, and profile-likelihood intervals. Their
software interface is described in Section~\ref{sec:uq_nl}, and the Bayesian
counterpart based on posterior samples is deferred to
Section~\ref{sec:bayes}.

\subsubsection{Wald-type and misspecification-robust intervals.}
Wald-type intervals rely on the asymptotic normality of maximum-likelihood estimators
\citep{vaart1998asymptotic} and use the observed information matrix, the negative Hessian of the
marginal log-likelihood of Equation~\ref{eq:mll} at the estimate, computed for the Laplace and
\gls{focei} estimators from the approximated marginal log-likelihood of Equation~\ref{eq:laplace} and
likewise for \gls{mcem} and \gls{saem}. Because the inverse observed information is valid only under correct model specification, the sandwich estimator \citep{white1982maximum} incorporates the empirical covariance of the individual score contributions to provide misspecification-robust standard errors. When the information-matrix equality holds, it coincides with the inverse observed information matrix.

\subsubsection{Profile-likelihood intervals.}
Profile-likelihood intervals \citep{meeker1995teaching} avoid the
normality assumption underlying Wald intervals by inverting the
likelihood-ratio statistic of each parameter against a $\chi^2$ threshold
\citep{raue2009structural}. They are particularly useful for parameters
with asymmetric uncertainty or estimates near the boundary of the
parameter space, and they can reveal practical non-identifiability that
Wald intervals often obscure. Their computation requires repeated
constrained optimizations and can therefore become demanding for large
models.

\subsection{Prediction}\label{sec:prediction}
Beyond estimating the population parameters and their uncertainty, a fitted model is used to predict responses, which in a mixed-effects model depends on whether the individual was
observed during model fitting. For an individual included in the training
data, the observed responses inform the posterior of the random effect in
Equation~\ref{eq:re_post}, and a prediction at a covariate value or time
point either substitutes the corresponding \gls{ebe} into the conditional
mean or averages the conditional mean over this posterior. For an
individual not observed during fitting, no such posterior is available, and
the prediction either plugs the covariate-dependent prior mean of the
random effect into the conditional mean or integrates the conditional mean
over the prior, the latter giving the population mean implied by the
marginal model of Equation~\ref{eq:marginal}. Within each setting the two
predictors coincide when the conditional mean is linear in the random
effect and generally differ otherwise \citep{lindstrom1990nonlinear,
lavielle2014mixed}. The corresponding expressions are given in
Appendix~\ref{app:prediction}.
In \nl{}, predictions for both observed and new individuals are available and
supply the values used by its diagnostics and plots (Section~\ref{sec:uq_nl}).

\subsection{Model evaluation}\label{sec:evaluation}

Because such predictions are only as trustworthy as the fit that produces them, a fitted model should be assessed for its absolute goodness of fit, the
adequacy of the assumed random-effects distribution, its performance
relative to competing specifications, and the reliability of both the
individual estimates and the fit itself. In \nl{}, we provide diagnostics and cross-validation for each of these aspects
(Section~\ref{sec:uq_nl}).

\subsubsection{Goodness-of-fit diagnostics.}
The adequacy of the fitted model is assessed by comparing its predictions
with the observed data. A \gls{vpc} compares observed data quantiles with prediction intervals obtained from replicate datasets simulated under the fitted model \citep{bergstrand2011prediction}. Residual diagnostics compare each observation with its population and individual predictions (\gls{pred} and \gls{ipred}, Equations~\ref{eq:pred_plugin} and~\ref{eq:pred_ebe}). The weighted residual (\gls{wres}) \citep{hooker2007conditional} standardizes the difference between an observation and its \gls{pred} by the corresponding predictive standard deviation. The \gls{pit} evaluates the fitted predictive cumulative distribution function at the observed value. Under correct model specification, the resulting values are uniformly distributed, and their standard-normal quantiles define normally distributed quantile residuals. 

\subsubsection{Random-effects diagnostics.}
In addition to reproducing the observed responses, a mixed-effects model
must adequately characterize between-individual variability through its
random-effects distribution. This assumption is assessed using diagnostics
based on the estimated random effects. The \gls{pit} of
the \glspl{ebe} under their fitted distribution is uniform when the random efefcts prior is correctly specified,
summarized as per-coordinate \gls{pit}.

Shrinkage quantifies the extent to which an \gls{ebe} is pulled toward the population mean when the data provide limited information about an individual. For each random-effect coordinate, the ETA shrinkage \citep{savic2009importance} is one minus the ratio of the empirical standard deviation of the \glspl{ebe}, $\mathrm{SD}(\hat{b})$, to the estimated random-effects standard deviation $\omega$, where $\hat{b}$ denotes the covariate-adjusted \gls{ebe}. Large shrinkage values indicate weak individual information and reduce the reliability of \gls{ebe}-based diagnostics. Values exceeding approximately $30\%$ are commonly considered problematic \citep{savic2009importance}.

\subsubsection{Model comparison and cross-validation.}
Competing model specifications are compared through their ability to
predict unseen data. In mixed-effects models, predictive performance
depends on whether predictions are made for previously observed
individuals or for entirely new individuals, as discussed in
Section~\ref{sec:prediction}. We therefore support both
observation-level and subject-level cross-validation, which target these
two prediction settings.

In subject-level cross-validation, entire individuals are assigned to
held-out folds. Because no observations from these individuals are
available during training, predictions are generated as for unseen
individuals, using either the plug-in predictor of
Equation~\ref{eq:pred_plugin} or the marginal predictor of
Equation~\ref{eq:pred_marginal}. This setting evaluates the ability of
the model to generalize to new individuals.

In observation-level cross-validation, only a subset of observations is
held out for each individual. The remaining observations are used to
estimate individual random effects, allowing predictions for the held-out
observations to be generated as for known individuals, using either the
\gls{ebe}-based predictor of Equation~\ref{eq:pred_ebe} or the posterior
average of Equation~\ref{eq:pred_posterior}. This setting evaluates the
ability of the model to predict additional observations from partially
observed individuals.

Performance is assessed using the held-out predictive log-likelihood or a
user-defined loss function.

\subsection{Extension to Bayesian hierarchical models}\label{sec:bayes}

The estimation and uncertainty quantification developed so far treat the population parameters $\theta$ as fixed unknowns and rely on large-sample approximations. The Bayesian
formulation equips $\theta$ with a prior $p(\theta)$, and targets the posterior of population and random effects parameters jointly \citep{gelman2013bayesian}.

\subsubsection{Bayesian model framework.}
The prior extends the two-level hierarchy of Equations~\ref{eq:nlme_re} and \ref{eq:nlme_obs} to a
three-level hierarchy in which the prior generates the population parameters, the population parameters
generate the random effects, and the random effects generate the observations
\citep{wakefield1996bayesian, lunn2002bayesian, gelman1995bayesian, davidian2003nonlinear}. Inference
targets the joint posterior of the population and random effects parameters,
\begin{equation}
  p(\theta, \{B_i\} \mid \{(y_i, x_i)\} )
  \propto
  p(\theta)
  \prod_{i=1}^N
  p(B_i \mid x_i, \theta)
  p(y_i \mid x_i, B_i, \theta),
  \label{eq:posterior}
\end{equation}
circumventing the need to estimate the integral in Equation~\ref{eq:marginal}, but replacing it with approximating a higher-dimensional posterior distribution, usually by \gls{mcmc}. 
The Bayesian formulation does not remove the latent random effects from the model. Rather, it treats them as unknown quantities to be inferred jointly with the population-level parameters.
In \nl{}, the same model specification is reused for Bayesian inference:
assigning priors to the population parameters switches estimation to \gls{mcmc}
sampling of the joint posterior of Equation~\ref{eq:posterior}
(Section~\ref{sec:est_nl}).

A third route is variational inference, which approximates the posterior by the closest member of a tractable family of distributions, found by maximizing the \gls{elbo}~\citep{blei2017variational}.

\subsubsection{Uncertainty quantification.}
Posterior uncertainty is summarized directly from the \gls{mcmc} draws that approximate the posterior of Equation~\ref{eq:posterior}. A credible interval for a
component $\theta_k$ is reported as a pair of posterior quantiles, in contrast to the Wald and
profile-likelihood intervals of Section~\ref{sec:uq}, and propagates parameter uncertainty without the
asymptotic-normality assumption \citep{gelfand1990sampling}. 

\section[Creating a NoLimits.jl model]{Creating a \nl{} model}\label{sec:model_macro}

A \nl{} analysis is organized around a small set of macros and data structures rather than a fixed sequence of function calls. The workflow consists of four steps. First, a \code{Model} is defined using the \code{@Model} macro independently of any dataset. Second, the model is combined with a dataset to form a \code{DataModel}. Third, the model is fit with \code{fit\_model}, producing a \code{FitResult} (Section~\ref{sec:est_nl}). Finally, uncertainty quantification and model diagnostics are performed using \code{compute\_uq} and the associated plotting utilities (Section~\ref{sec:uq_nl}).

The macro-based design provides the expressiveness of a high-level modeling language, while \proglang{Julia}'s multiple dispatch enables estimators, uncertainty-quantification methods, and model components to remain interchangeable and user-extensible. The remainder of this section introduces the \code{@Model} macro by constructing a model specification step by step, covering fixed effects, covariates, random effects, structural dynamics, and observation models.

\subsection{The model macro}

The first of the four steps, defining a \code{Model}, is carried out entirely with the \code{@Model} macro, which assembles the building blocks summarized in Table~\ref{tab:blocks}.

Each block is defined through a dedicated sub-macro, of which only \code{@fixedEffects} and \code{@formulas} are required. The blocks may be specified in any order and are sorted automatically during model construction.

As a minimal example, an exponential-decay model with a subject-specific decay rate can be written as follows.

\begin{CodeInput}
model = @Model begin
    @fixedEffects begin
        A     = RealNumber(10.0, scale = :log)
        k_pop = RealNumber(0.0)
        omega = RealNumber(0.3, scale = :log)
        sigma = RealNumber(0.2, scale = :log)
    end
    @covariates begin
        t = Covariate()
    end
    @randomEffects begin
        k = RandomEffect(LogNormal(k_pop, omega); column = :id)
    end
    @formulas begin
        mu = A * exp(-k * t)
        y ~ Normal(mu, sigma)
    end
end
\end{CodeInput}

While intentionally simple, this example illustrates the core structure of a model specification. In practice, the same blocks can express substantially richer models, while additional macros listed in Table~\ref{tab:blocks} provide support for specialized components such as \glspl{ode}.

\begin{table}[!ht]
  \centering
  \begin{tabularx}{\linewidth}{@{}llXc@{}}
    \toprule
    Block & Macro & Role & Required \\
    \midrule
    Helpers        & \code{@helpers}                 & Reusable sub-expressions shared across blocks           &              \\
    Fixed effects  & \code{@fixedEffects}            & Population parameters with scales, bounds, and priors    &   $\checkmark$           \\
    Covariates     & \code{@covariates}              & Time-varying, group-constant, and dynamic covariates     &              \\
    Random effects & \code{@randomEffects}           & Random effects and their distributions                   &              \\
    Pre-equation   & \code{@preDifferentialEquation} & Quantities derived before solving the dynamics           &              \\
    Dynamics       & \code{@DifferentialEquation}    & The system of ordinary differential equations            &              \\
    Initial state  & \code{@initialDE}               & Initial conditions of the dynamics                       &              \\
    Observations   & \code{@formulas}                & Deterministic signals and the observation distribution   & $\checkmark$ \\
    \bottomrule
  \end{tabularx}
  \caption{Building blocks of a \code{@Model} specification, in the canonical
    order in which the macro arranges them.}
  \label{tab:blocks}
\end{table}

\subsection{Fixed effects}
The \code{@fixedEffects} block declares the population parameters and, where supported by the estimation method, the scale on which they are estimated. Each line defines a named parameter, whose identifier can subsequently be referenced throughout the model specification. To accommodate the diverse parameter types introduced in Section~\ref{sec:nlme}, \nl{} provides parameter blocks ranging from unconstrained scalars to structured and constrained matrices. Table~\ref{tab:paramtypes} summarizes the available parameter types and their associated reparameterizations.

\begin{table}[!ht]
  \centering
  \begin{tabular}{lll}
    \toprule
    Type & Represents & Scales \\
    \midrule
    \multicolumn{3}{l}{\emph{Scalar, vector, and matrix parameters}} \\
    \code{RealNumber}                 & Scalar                  & \code{:identity}, \code{:log}, \code{:logit} \\
    \code{RealVector}                 & Vector                  & \code{:identity}, \code{:log}, \code{:logit} \\
    \code{RealDiagonalMatrix}         & Diagonal matrix         & \code{:log} \\
    \code{RealPSDMatrix}              & Covariance matrix       & \code{:cholesky}, \code{:expm} \\
    \code{ProbabilityVector}          & Probability vector      & \code{:stickbreak} \\
    \code{DiscreteTransitionMatrix}   & Transition matrix       & \code{:stickbreakrows} \\
    \code{ContinuousTransitionMatrix} & Rate matrix             & \code{:lograterows} \\
    \addlinespace
    \multicolumn{3}{l}{\emph{Differentiable machine-learning components}} \\
    \code{NNParameters}               & Neural network          & -- \\
    \code{SoftTreeParameters}         & Soft decision tree      & -- \\
    \code{SplineParameters}           & Spline basis            & -- \\
    \code{NPFParameter}               & Normalizing planar flow & -- \\
    \bottomrule
  \end{tabular}
    \caption{Fixed-effect parameter types, what each represents, and the
    reparameterizations available through the \code{scale} keyword. The scale
    defaults to the first one listed for each type, and the matrix, probability,
    and transition scales keep each parameter in its valid set, positive-definite,
    on the simplex, or row-stochastic. The differentiable machine-learning components take no
    \code{scale} argument and are estimated on an unconstrained basis.}
  \label{tab:paramtypes}
\end{table}

\subsubsection{Values, scales, bounds, and priors.}
The scalar, vector, and matrix parameters (Table~\ref{tab:paramtypes}) are initialized from a starting value passed as the first
positional argument, as in \code{RealNumber(0.0)}, with the remaining arguments optional keywords.
The \code{scale} keyword selects the reparameterization from Table~\ref{tab:paramtypes} and so
determines the unconstrained coordinates used by the optimizer (Section~\ref{sec:est_nl}): a positive
parameter uses \code{scale = :log} and a probability \code{scale = :logit}, and a vector may carry a
separate scale per element. The \code{lower} and \code{upper} keywords add box constraints for scalar
and vector parameters. The \code{prior} keyword attaches a \pkg{Distributions.jl} distribution used
by the Bayesian estimators and ignored by the frequentist ones, and
\code{calculate\_se} controls whether a parameter enters the standard-error computation, defaulting to
off for covariance matrices and differentiable machine-learning components.

\subsubsection{Differentiable machine-learning components.}
The differentiable machine-learning types of Table~\ref{tab:paramtypes} are also declared in the \code{fixedEffects}
but follow a different convention. Rather than a starting value and scale, each
takes the structure that defines it, such as the network passed to
\code{NNParameters}, together with a \code{function\_name} used to call it from
other blocks, and its coefficients are initialized internally.
Section~\ref{sec:learned} outlines how they are embedded in a model. 

\subsection{Helpers}

Model blocks frequently share the same algebraic expressions. The \code{@helpers} block defines small utility functions, such as link or saturation functions, that are reused across them. These functions could equally be defined in the surrounding \proglang{Julia} scope, but declaring them inside \code{@helpers} attaches them directly to the resulting \code{Model}, so that the model carries its own helpers and remains self-contained. Each entry is an ordinary function definition, for example \code{softplus(u) = log1p(exp(u))}, and is afterwards available by name in \code{@randomEffects}, \code{@preDifferentialEquation}, \code{@DifferentialEquation}, \code{@initialDE}, and \code{@formulas}. The block is optional and keeps repeated algebra in one place.

\subsection{Covariates}

Whereas fixed effects and helpers are internal to the model, covariates link it to the dataset. The \code{@covariates} block declares which columns are used by the model and how they are accessed during evaluation. Each covariate is assigned a name and a covariate type, for example \code{t = Covariate()}. The declared name is subsequently available throughout the model specification and is also used to retrieve values from the dataset. Consequently, each covariate name must correspond to a column in the data frame supplied when fitting the model.

\nl{} distinguishes between three covariate types, each of which also has a vector-valued variant. The covariate type determines where in the model the covariate may be used and how its values are accessed.

\subsubsection{Covariate types.}

1) A \code{Covariate()} is evaluated row by row and is therefore available only within the observation model specified in \code{@formulas}. 2) A \code{ConstantCovariate(; constant\_on = :id)} is assumed constant within each individual and may be used throughout the model specification, allowing random-effects distributions, \gls{ode} initial conditions, \gls{ode} dynamics, and other model components to depend on observed covariates. 3) A \code{DynamicCovariate(; interpolation = LinearInterpolation)} is converted into a continuous function of time and can therefore be evaluated at arbitrary time points within the \code{@DifferentialEquation} block. For example, a dynamic covariate declared as \code{rate} is accessed as \code{rate(t)}. The interpolation method is selected from \pkg{DataInterpolations.jl} \citep{Bhagavan2024}, enabling piecewise-constant, linear, and nonlinear interpolation schemes.

Each covariate type also has a vector-valued counterpart (\code{CovariateVector} and its constant and dynamic variants), which groups multiple columns under a single name. The grouped name then behaves directly as a numeric vector, so it can be passed as a whole wherever a vector is expected, while the individual columns remain accessible by name. This is particularly useful when covariates are used collectively, for example as inputs to a neural network, as in \code{NN(x, nn)}.

\subsection{Random effects}

The \code{@randomEffects} block defines the between-subject variability of the model through the random effects $B_i$ introduced in Section~\ref{sec:nlme}. Each random effect is associated with a grouping variable and is shared by all observations within the same group. As with covariates, the grouping variable must correspond to a column in the dataset supplied during model fitting.

\subsubsection{Declaring random effects.}

A random effect is declared as \code{eta = RandomEffect(dist; column = ...)}, where \code{eta} is the name of the random effect, \code{dist} specifies its distribution, and \code{column} identifies the grouping variable. Distribution parameters may depend on fixed effects and constant covariates, allowing covariate-dependent random-effects distributions. Multiple random effects can be defined on the same or different grouping variables, supporting nested and crossed random-effects structures (Appendix~\ref{app:nested}).

Once declared, a random effect can be referenced by name in all model blocks listed in Table~\ref{tab:blocks}, except \code{@fixedEffects}.

\subsubsection{Flexible random-effects distributions.}
Random effects can be modeled using arbitrary continuous distributions with finite mean from \pkg{Distributions.jl}, including heavy-tailed, skewed, strictly positive, and multivariate distributions such as \code{MvNormal}. Components of multivariate random effects can be accessed by indexing, for example \code{eta[1]}. 

When a modeler aims for a nonparametric random effects distribution, the random-effects distribution can instead be learned from the data using a normalizing planar flow. For example, a flow parameter declared among the fixed effects as \code{psi = NPFParameter(2, 2)} defines a random effect through \code{RandomEffect(NormalizingPlanarFlow(psi); column = :id)}. The flow parameters are estimated jointly with the remaining model parameters, enabling flexible multivariate distributions that can capture skewness and multimodality through \pkg{NormalizingFlows.jl}.

\subsection{Structural dynamics}

The blocks introduced so far describe quantities evaluated directly from parameters and covariates. Mechanistic models additionally evolve internal states over time, specified through up to three \gls{ode} blocks: one for quantities computed before integration, one for the system dynamics, and one for the initial conditions. All three blocks have access to fixed effects, random effects, constant covariates, and arbitrary transformations thereof, while the \code{@DifferentialEquation} block may additionally access dynamic covariates.

The following example illustrates the basic structure,

\begin{CodeInput}
@covariates begin
    dose = ConstantCovariate()
end
@preDifferentialEquation begin
    k = exp(log_k + eta_k)
end
@DifferentialEquation begin
    D(A) ~ -k * A
    flux(t) = k * A
end
@initialDE begin
    A = dose
end
\end{CodeInput}
which we further explain below. 

\subsubsection{Pre-integration computations.}

The optional \code{@preDifferentialEquation} block is used for calculations that do not depend on the state trajectory and therefore need only be performed once per individual. Typical examples include transforming fixed and random effects into individual parameters, as in \code{k = exp(log_k + eta_k)}. Although such expressions could be placed directly in the differential equations, evaluating them before integration avoids unnecessary computations during solver execution.

\subsubsection{Dynamical system.}

The \code{@DifferentialEquation} block defines the system dynamics through equations of the form \code{D(state) ~ rhs}. It may additionally define named signals such as \code{flux(t) = k * A}, representing derived quantities that depend on time and the current state. Both states and signals are available to the observation model and may be referenced elsewhere in the model specification.

The resulting system is solved using \pkg{OrdinaryDiffEq.jl}, and the solution is exposed as a continuous function of time.

\subsubsection{Accessing states and signals.}

States and signals can be accessed in the \code{@formulas} block as \code{A(t)} and \code{flux(t)}. Because the solution is continuous in time, they may also be evaluated at shifted time points, such as \code{A(t - 0.5)}, enabling the representation of delays and carry-over effects.

\subsubsection{Initial conditions.}

Initial conditions are specified in the \code{@initialDE} block. In the example above, the initial state depends on the covariate \code{dose}. More generally, initial conditions may depend on fixed effects, random effects, covariates, or learned model components. 

\subsection{Observation models}

The \code{@formulas} block is where states, fixed and random effects,
and covariates are linked to the conditional observation models
(Equation~\ref{eq:nlme_obs}). It is the most flexible block in the model
definition and has access to all previously defined quantities.

We use the following example to illustrate its capabilities.

\begin{CodeInput}
@formulas begin
    conc  = A(t) / V
    c     ~ Normal(conc, sqrt(sigma_add^2 + (sigma_prop * conc)^2))
    count ~ Poisson(exp(eta + beta * t))
end
\end{CodeInput}

The block distinguishes between two types of specifications. The first
defines a deterministic signal, such as \code{conc = A(t) / V}, using an
equality sign. The second defines a distributional relationship, such as
\code{c ~ Normal(conc, sqrt(sigma\_add\textasciicircum{}2 + (sigma\_prop * conc)\textasciicircum{}2))},
using the \code{\textasciitilde{}} operator.

\subsubsection{Deterministic signals.}

A deterministic line of the form \code{name = term} introduces a named
signal. This signal can be further processed within
the \code{@formulas} block or passed to a distributional relationship to
parameterize a conditional observation distribution. For example,
\code{conc = A(t) / V} defines the signal \code{conc}, which is
subsequently used in the observation model.

The defining expression can be nearly arbitrarily complex and may involve any
previously defined quantities, including states, parameters, fixed and
random effects, and covariates.

\subsubsection{Distributional relationships.}

A line of the form \code{name\_y ~ dist} links an observed variable to a
conditional probability distribution, as in
\code{name\_y ~ Poisson(exp(eta + beta * t))}. The variable
\code{name\_y} must correspond to a column in the observation data and
is automatically associated with the specified conditional distribution.

Any univariate distribution from \pkg{Distributions.jl} that admits a
finite density under the chosen parameterization can be used. As a
result, continuous, count, binary, and other observation types are
specified uniformly. Vector-valued distributions can additionally be
used to model correlation within the conditional observation model.

Because the parameters of a distribution are themselves arbitrary
expressions, complex dependencies between the conditional observation
distribution and other model quantities can be specified directly.

For example, a constant standard deviation,
\code{c ~ Normal(conc, sigma)}, defines an additive error model,
whereas a standard deviation proportional to the prediction,
\code{c ~ Normal(conc, sigma\_prop * conc)}, defines a proportional
error model. Combining additive and proportional terms yields a
combined additive-and-proportional error model.

\subsubsection{Multiple observations.}

Multiple observation models can be specified within the same
\code{@formulas} block when multiple observed variables are available.
In the example above, \code{c} is a continuous variable and
\code{count} is a count variable. Consequently, each individual
contributes observations from two different distribution families
(Appendix~\ref{app:multi}).

\subsubsection{Censored observations.}

Censored observations, whose values are only known to lie within an
interval (for example, below a lower limit of quantification), are
supported through the \pkg{Distributions.jl} interface. One wraps the
observation distribution as

\code{c ~ censored(Normal(conc, sigma), lower = lloq, upper = Inf)},

where either bound may be specified as a constant value shared across
all observations or as a covariate column, allowing observation-specific
censoring limits (Appendix~\ref{app:censor}). Combining a survival-time
observation distribution such as \code{Exponential} or \code{Weibull} with
right-censoring in this way expresses parametric time-to-event models, which
therefore require no dedicated machinery beyond the general observation-model
and censoring interface.

\subsubsection{Markov observation models.}
In \nl{}, Markov models are implemented as observation distributions and
are therefore specified in the same way as any other stochastic model
component. \nl{} provides hidden- and observed-state models in discrete-
and continuous-time variants, with univariate emissions, multivariate
emissions with missing observations, and coarsened observed states
(Table~\ref{tab:markov}). All components of these models can be defined as
arbitrary expressions and may therefore depend on covariates, fixed and
random effects, or learned components such as neural networks. A worked
\gls{hmm} specification, together with the forward recursion that defines
its conditional likelihood, is given in Appendix~\ref{app:hmm}.

\subsection{Embedding differentiable machine-learning components}\label{sec:learned}

A distinguishing feature of \nl{} is that differentiable machine
learning components are treated as ordinary parameter blocks and can be
used throughout a model. The available component types are listed in
Table~\ref{tab:paramtypes}. Each component is declared in
\code{@fixedEffects} via
\code{params = NNParameters(args; function_name = :func\_name)}, where
the function name is mandatory. The component can then be called from
any other block as \code{func\_name(inputs, params)}, passing the input
vector and the parameter block.

Neural networks and soft decision trees return vectors and therefore
require indexing to select a particular output component, for example
\code{func\_name(inputs, params)[j]}. In contrast, splines map a scalar
input to a scalar output and can be used directly without indexing. The
example below declares one component of each type and uses the neural
network in the random-effect distribution, the soft tree in the system
dynamics, and the spline in the observation model. The model serves only
to illustrate the flexibility with which machine-learning components can
be integrated into \nl{} models.
  \begin{CodeInput}
  using SimpleChains
  chain = SimpleChain(static(2), TurboDense(tanh, 4), TurboDense(identity, 1))

  @fixedEffects begin
      theta     = RealVector([0.5, 1.0, 0.3, 0.2], scale = fill(:log, 4))
      par_nn    = NNParameters(chain; function_name = :NN)
      par_st    = SoftTreeParameters(2, 3; function_name = :ST)
      par_sp    = SplineParameters(0.0:1.0:10.0; function_name = :SP)
  end
  @covariates begin
      x = ConstantCovariateVector([:age, :bmi]; constant_on = :id)
  end
  @randomEffects begin
      eta = RandomEffect(
               Normal(NN(x, par_nn)[1], theta[3]); column = :id
            )
  end
  @DifferentialEquation begin
      D(A) ~ ST([A, t], par_st)[1] - theta[1] * A
  end
  @formulas begin
      conc = A / theta[2]
      y ~ Normal(SP(conc, par_sp) + eta, theta[4])
  end
  \end{CodeInput}
In \code{@randomEffects}, the network \code{NN} maps the subject-level
covariates \code{age} and \code{bmi} to the mean of the random effect
distribution. Because random effects are defined once per subject, their
inputs must be constant within a subject. In
\code{@DifferentialEquation}, the soft tree \code{ST} adds a learned
state- and time-dependent term to the dynamics of \code{A}, turning the
system into a \gls{ude}. Finally, in \code{@formulas}, the spline
\code{SP} maps the deterministic signal \code{conc} to the conditional
observation mean, with its coefficients estimated as fixed effects on
the knot vector specified in \code{SplineParameters}.

Because machine-learning components are ordinary parameter blocks, new
component types can be integrated by defining their parameter type and
evaluation function, without modifying the rest of the framework.

The network passed to \code{NNParameters} is a
\pkg{SimpleChains.jl} \code{SimpleChain} with two inputs, a hidden layer
of four \code{tanh} units, and a linear output layer. The same code
would also work with a \pkg{Lux.jl} \code{Chain}, with only the network
object changing. The call convention, estimators, and optimization
routines remain unchanged. We generally recommend
\pkg{SimpleChains.jl} for the small CPU-resident networks commonly used
in \gls{nlme} models because it is faster and more memory efficient,
while \pkg{Lux.jl} is preferable when more expressive architectures are
required.

\subsection{Binding a model to data}
A \code{Model} is independent of any particular dataset and is bound to one with
the \code{DataModel} constructor.

\subsubsection{Data format.}
The data are supplied as an ordinary \pkg{DataFrames.jl} \citep{bouchetvalat2023dataframes}
data frame in long format, passed as
\code{df} in \code{DataModel(model, df; primary\_id = :id, time\_col = :t)}, with
\code{primary\_id} the subject identifier and \code{time\_col} the time column. The
data frame stacks the records of all subjects on top of one another, one row per
record, the layout used by \pkg{NONMEM} and \pkg{Monolix}. Each row carries the
subject identifier, the time, the observations named on the stochastic lines of
\code{@formulas}, the covariates named in \code{@covariates}, and the grouping
columns named in \code{@randomEffects}. An observation that is not measured in a given
row is left \code{missing}, so a row that records only a dose or a covariate value
carries \code{missing} in its observation columns.

\subsubsection{Dosing and events.}
For \gls{ode}-based models, doses and other interventions are specified
through event records in the data, following the conventions of
\pkg{NONMEM} and \pkg{Monolix}. Event handling is enabled by passing an
event column to the data model \code{DataModel} construction,
\begin{CodeInput}
dm = DataModel(model, df;
    primary_id = :id,
    time_col = :t,
    evid_col = :EVID)
\end{CodeInput}
Rows with \code{EVID = 0}, \code{1}, and \code{2} denote observations,
doses, and compartment resets, respectively. Doses may be administered
as boluses or infusions, and target compartments are specified through \code{CMT} (Table~\ref{tab:events}). 

\begin{table}[t!]
  \centering
  \begin{tabularx}{\linewidth}{@{}llX@{}}
    \toprule
    Column & Keyword & Role \\
    \midrule
    \code{EVID} & \code{evid\_col} & Event type, with \code{0} an observation, \code{1} a dose, and \code{2} a compartment reset \\
    \code{AMT}  & \code{amt\_col}  & Dose amount, or the value a compartment is set to on a reset \\
    \code{RATE} & \code{rate\_col} & Infusion rate, with \code{0} a bolus and for a positive value an infusion \\
    \code{CMT}  & \code{cmt\_col}  & Target compartment, given as a state name \\
    \bottomrule
  \end{tabularx}
  \caption{Event columns of the long data table, read when \code{evid\_col} is
    set. The keyword column gives the \code{DataModel} argument that names each
    column, and the amount, rate, and compartment keywords default to
    \code{:AMT}, \code{:RATE}, and \code{:CMT}.}
  \label{tab:events}
\end{table}

\subsubsection{Construction and validation.}
Constructing the
\code{DataModel} checks that the required columns are present, partitions the
rows by the primary identifier into one \code{Individual} per subject, and
precomputes the grouping structure used by the random-effect blocks. The
resulting \code{DataModel} is the object passed to every estimator in the next
section.

\section[Parameter estimation in NoLimits.jl]{Parameter estimation in \nl{}}\label{sec:est_nl}

All estimation methods in \nl{} share a common interface. Each estimator
is represented by a Julia struct, and a model is fit by passing a
\code{DataModel} together with the desired estimator to
\code{fit\_model}. Consequently, switching between estimation methods
requires changing only the estimator struct, while the model
specification and data binding remain unchanged.
Estimation with the Laplace method only requires only minimal code.
\begin{CodeInput}
res_lap  = fit_model(dm, Laplace()) 
\end{CodeInput}
Other estimators can be used by replacing \code{Laplace()}, with the corresponding estimator constructor.
Table~\ref{tab:estimators} summarizes the available estimators, the
inference paradigms they implement, their support for random effects,
and their typical use cases. Together, they cover the methods described
in Sections~\ref{sec:point_est} and~\ref{sec:bayes}. If an estimator is
incompatible with the specified model, for example because a
fixed-effects-only method is applied to a mixed-effects model or a
Bayesian estimator is used without prior distributions, \nl{} produces
an informative error message. Variational inference is currently available for
fixed-effects-only models through the \code{VI} estimator, with variational
inference over the random effects planned as a future extension.

All estimation methods share the numerical optimization, parameter
transformation, and multistart machinery described in the remainder of
this section. Mixed-effects models can additionally be warm-started from
a pooled fit that ignores the random-effect structure by passing
\code{pooled\_init = true} to \code{fit\_model}.

\begin{table}[t!]
  \centering
  \begin{tabularx}{\linewidth}{@{}llcX@{}}
    \toprule
    Estimator & Inference & Random or Fixed & Typical use \\
    \midrule
    \code{MLE}        & Likelihood             & Fixed      & Models without random effects \\
    \code{MAP}        & Posterior   & Fixed      & Fixed effects with priors \\
    \code{Pooled}     & Likelihood             & Fixed & Starting values for mixed-effects fits \\
    \code{PooledMap}     & Posterior             & Fixed & Starting values for mixed-effects fits with priors \\
    \code{Laplace}    & Approximate likelihood & Random     & Small number of random effects\\
    \code{FOCEI}      & Approximate likelihood & Random     & Moderate number of random effects \\
    \code{MCEM}       & Stochastic \acrshort{em} & Random   & When the Laplace approximation is inadequate \\
    \code{SAEM}       & Stochastic \acrshort{em} & Random   & Moderate number of random effects and/or non-Gaussian posterior \\
    \code{MCMC}       & Bayesian sampling            & Both    & Full posterior inference \\
    \code{VI}         & Approximate Bayesian   & Fixed      & Fast approximate posterior \\
    \bottomrule
  \end{tabularx}
  \caption{Estimators available through \code{fit\_model}, the inference
    paradigm each implements, whether it accommodates random effects, and the
    setting in which each is typically used.}
  \label{tab:estimators}
\end{table}

\subsection{Numerical optimization and parameter transformations}

The estimators in \nl{} share a set of numerical components that are configured separately from the model and its data binding. Each can be swapped or
returned on its own, leaving the model specification and the remaining components untouched.

\subsubsection{Numerical optimization.}
Whenever numerical optimization is required, \nl{} interfaces with
\pkg{Optimization.jl}, providing access to a broad range of gradient-based
and gradient-free optimization algorithms from packages such as
\pkg{Optim.jl} \citep{mogensen2018optim}, \pkg{NLopt.jl} \citep{NLopt} and \pkg{BlackBoxOptim.jl} \citep{blackboxoptimjl}. The
default optimizer for the Laplace and \gls{focei} outer optimization is the gradient-free BOBYQA \citep{powell2009bobyqa} algorithm implemented in NLopt.jl. For all other numerical optimization tasks, the default is the \gls{lbfgs} algorithm \citep{liu1989limited}
and gradients are computed via
forward-mode automatic differentiation using \pkg{ForwardDiff.jl}. Every estimator that performs numerical optimization exposes the same two controls, the
\code{optimizer} object and the \code{optim\_kwargs} named tuple, the latter
forwarded verbatim to \code{Optimization.solve} so that iteration caps and
solver tolerances are set without leaving the estimator constructor.
Selecting a different algorithm is a one-line change that leaves the model
and data specification untouched. For example,
\begin{CodeInput}
res = fit_model(dm, Laplace(optimizer = NoLimits.LBFGS(),
                            optim_kwargs = (maxiters = 200,)))
\end{CodeInput}
will use the \gls{lbfgs} implementation from \pkg{Optim.jl} for the outer optimization of the Laplace algorithm. 
The automatic differentiation backend is currently
fixed to forward mode.

\subsubsection{Parameter transformations.}
Parameter transformations are defined by the \code{scale} keyword argument when defining fixed effects (Table~\ref{tab:paramtypes}). The scale transforms a constrained parameter, e.g., a positive real number, to an unconstrained parameter allowing the optimizer to operate in an unconstrained space. The inverse
map recovers the natural values for likelihood evaluation. More complex examples include parameterizations of variance covariance matrices, which must be positive definite. This is achieved in \nl{} by estimating the Cholesky factorization (\code{scale = :cholesky}) or its matrix logarithm and matrix exponential (\code{scale = :expm}).
Domains are
therefore enforced by construction rather than through box constraints or penalty
terms, so the optimization stays unconstrained even when the parameters are
bounded, positive-definite, or confined to a simplex. This is in contrast to specifying box-constraints via \code{lower} and \code{upper}, which uses constrained optimization.

For Bayesian estimation, priors are specified on the untransformed parameter
scale. Consequently, Bayesian methods do not use the parameter
transformations or box constraints specified for frequentist estimation.
Parameter values outside the admissible domain pf the prior are instead assigned an
objective value of $-\infty$.

Two further numerical components are configured in the same separable way,
the \code{Multistart} procedure that guards against local optima and the
\gls{ode} solver selected. Both are
described in Appendix~\ref{app:config}.

\section[Uncertainty quantification and visualization in NoLimits.jl]{Uncertainty quantification and visualization in \nl{}}\label{sec:uq_nl}

Uncertainty quantification in \nl{} is performed through the single
function \code{compute\_uq}, which takes a \code{FitResult} and returns
a \code{UQResult}. Depending on the chosen method, the result contains
point estimates together with confidence or credible intervals and,
where available, the covariance matrix and the underlying parameter
draws. All quantities are reported on both the natural and transformed
parameter scales.

The uncertainty construction is selected through the
\code{method} keyword and its coverage through \code{level}. By default,
\nl{} chooses a method consistent with the estimator used for fitting,
for example Wald intervals for frequentist estimators and
posterior intervals for Bayesian estimators. These constructions
correspond to the methods described in Section~\ref{sec:uq}.

A \code{UQResult} can be inspected with \code{summarize(uq)} and further
visualized with the provided function \code{plot\_uq\_distributions}. The remainder of this
section introduces the plotting functions used to assess fitted models,
their predictions, and the associated uncertainty.

\subsection{Wald and sandwich intervals}
With \code{method = :wald}, confidence intervals are computed from the
inverse observed information matrix, while
\code{method = :sandwich} uses the robust sandwich covariance estimator
(Section~\ref{sec:uq}). 
\begin{CodeInput}
uq = compute_uq(res; method = :wald, level = 0.95)
\end{CodeInput}
These methods are computationally inexpensive and
therefore the default for likelihood-based estimators. Their validity
relies on large-sample asymptotics and, in practice, on the objective
function being approximately quadratic in a neighborhood of the optimum.

\subsection{Profile-likelihood intervals}

With \code{method = :profile}, \nl{} computes profile-likelihood
intervals through \pkg{LikelihoodProfiler.jl}
\citep{borisov2026likelihoodprofiler}.
\begin{CodeInput}
uq = compute_uq(res; method = :profile)
\end{CodeInput}
Unlike Wald intervals, these do
not rely on a local quadratic approximation of the log-likelihood and are
therefore more reliable for asymmetric or weakly identifiable parameters,
at the cost of potentially costly repeated model refits (Section~\ref{sec:uq}).

\subsection{Posterior intervals}
With \code{method = :chain}, uncertainty intervals are obtained from
quantiles of posterior samples,
\begin{CodeInput}
uq = compute_uq(res; method = :chain)
\end{CodeInput}

For an \code{MCMC} fit, the posterior samples are already available. For
likelihood-based fits, \nl{} can optionally refit the model using
\gls{mcmc} and compute intervals from the resulting posterior draws.
These intervals do not rely on asymptotic approximations and correspond
to Bayesian credible intervals (Section~\ref{sec:bayes}). Users should be
aware that, in the latter case, uncertainty quantification is based on a
Bayesian rather than a frequentist interpretation of probability.

\subsection{Unified plotting style}
\nl{} provides a family of plotting functions, built on \pkg{Plots.jl} \citep{plotsjl}, that act
on a \code{FitResult}, or, for parameter uncertainty, on a \code{UQResult}.
They share a common \code{PlotStyle} options object, so a consistent appearance
is applied across an analysis by constructing one \code{PlotStyle} and passing
it to each call through the \code{style} keyword.

\begin{CodeInput}
ps = PlotStyle()
plot_vpc(res; style = ps)
plot_residual_qq(res; style = ps)
\end{CodeInput}

Each function returns a \pkg{Plots.jl} object that is displayed, saved, or
composed like any other plot. 

\section{Examples}\label{sec:examples}

We illustrate the main functionality of \pkg{NoLimits.jl} through three case studies
that span different model classes supported by the package: a compartmental
\acrshort{ode} model for warfarin pharmacokinetics
(Section~\ref{sec:warfarin}), a learnable concentration-effect function embedded in
the pharmacodynamic dynamics of the same drug
(Section~\ref{sec:warfarin_pkpd}), and a normalizing flow as a flexible
random-effects distribution (Section~\ref{sec:flow}). Further supported model
classes, including the discrete- and continuous-time Markov models described in
Section~\ref{sec:model_macro} and Appendix~\ref{app:hmm}, are documented in the
software sections rather than exercised in a worked example here. All scripts and
figure-rendering code are provided in the replication code. The code blocks
below show only the \pkg{NoLimits.jl} API calls that constitute the core analysis
workflow. We use a common \code{PlotStyle} type for all \nl{} plotting functions
for a unified plot styling.
All figures across the three case studies share a single instance:

\begin{CodeInput}
my_style = PlotStyle(
    base_subplot_width  = 500,
    base_subplot_height = 400,
    font_size_title      = 18,
    font_size_label      = 15,
    font_size_tick       = 14,
    font_size_legend     = 14,
    font_size_annotation = 12,
)
\end{CodeInput}

\subsection[Warfarin pharmacokinetics: a one-compartment ODE model]{Warfarin
  pharmacokinetics: a one-compartment \acrshort{ode} model}
\label{sec:warfarin}

Warfarin is an anticoagulant typically used to treat blood clots. It can exhibit substantial inter-individual variability in exposure,
making it a standard benchmark for population pharmacokinetic modeling
\citep{lane2012population}.
A common approach is a compartmental \gls{ode} embedded in an
\gls{nlme} model with random effects and weight-based covariates.
\nl{} provides a composable implementation of this model class with
interchangeable estimators and built-in diagnostics.
We fit a warfarin model using \gls{mcem}, quantify uncertainty, assess
fit, characterize inter-individual variability, and then refit the same model
by \gls{mcmc} to obtain a full Bayesian posterior.

\subsubsection{Data}
We use the warfarin pharmacokinetics dataset from the \pkg{Monolix}
tutorial library \citep{Monolix2024R1}. It contains repeated plasma
concentration measurements for 32 healthy volunteers following a single
oral dose. The dataset is loaded with
\code{load\_warfarin\_from\_monolix()}, which downloads the data and
converts them directly to the \nl{} format. In the analysis below, we use
the subject identifier (\code{id}), observation time (\code{t}), dose
(\code{d}), body weight (\code{wt}), and plasma concentration
(\code{C}).

\begin{CodeInput}
df = NoLimits.load_warfarin_from_monolix()
first(select(df, :id, :t, :d, :wt, :C), 5)
\end{CodeInput}

\begin{CodeOutput}
5x5 DataFrame
 Row | id      t        d        wt       C
     | String  Float64  Float64  Float64  Float64?
-----+----------------------------------------------
   1 | 1           0.0    100.0     66.7  missing
   2 | 1          24.0    100.0     66.7      9.2
   3 | 1          36.0    100.0     66.7      8.5
   4 | 1          48.0    100.0     66.7      6.4
   5 | 1          72.0    100.0     66.7      4.8
\end{CodeOutput}

\begin{figure}[t!]
  \centering
  \includegraphics[width = \textwidth]{%
    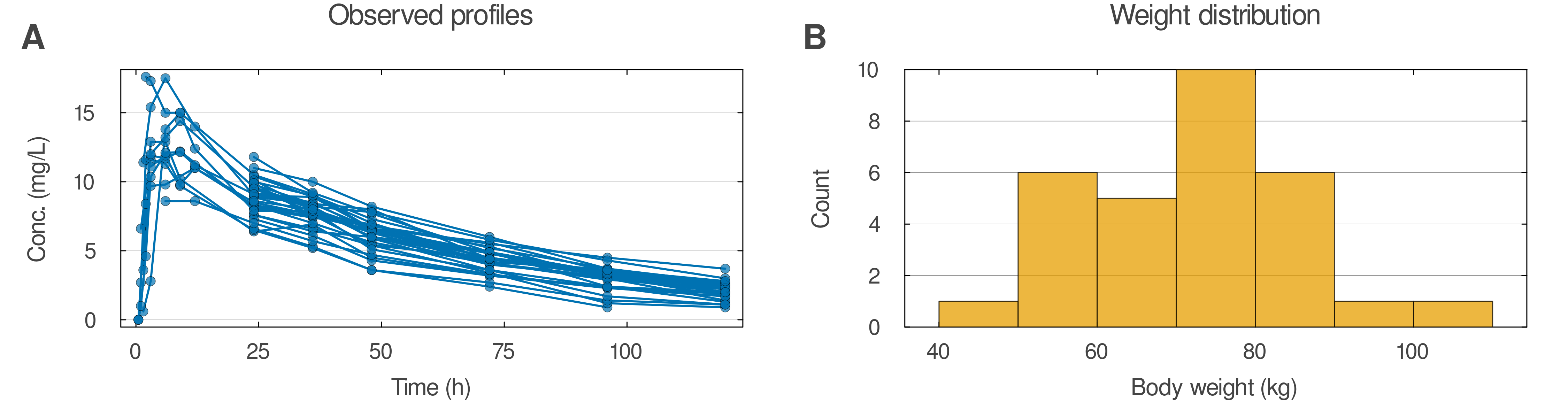}
  \caption{(A)~Observed plasma warfarin concentrations versus time for 32 subjects.
    Each line represents one subject. Produced with the \nl{} function
\code{plot\_observed\_profiles}. (B)~Body weight distribution across subjects.}
  \label{fig:warfarin_data}
\end{figure}

The first five rows show that the dose \code{d} and the weight \code{wt} are constant within a subject,
while time \code{t} and concentration \code{C} vary across measurement occasions.
The plasma concentration profiles exhibit the characteristic absorption peak followed by
log-linear elimination, with substantial inter-individual variability in both peak
concentration and elimination rate (Figure~\ref{fig:warfarin_data}A). Body weight ranges
from 40 to 102~kg (Figure~\ref{fig:warfarin_data}B).

\subsubsection{Model specification}

We specify a one-compartment first-order absorption pharmacokinetic model to describe the warfarin plasma concentration-time profiles.
For
subject $i$ receiving dose $d_i$, the depot amount $A_{\mathrm{dep},i}(t)$ and
central-compartment amount $A_{\mathrm{cen},i}(t)$ evolve according to the \gls{ode}
system
\begin{equation}
\begin{alignedat}{2}
\frac{dA_{\mathrm{dep},i}}{dt}
&= -k_{a,i}\,A_{\mathrm{dep},i}(t),
\qquad
&
\frac{dA_{\mathrm{cen},i}}{dt}
&= k_{a,i}\,A_{\mathrm{dep},i}(t)
   - \frac{\mathrm{CL}_i}{V_i}\,A_{\mathrm{cen},i}(t).
\end{alignedat}
\label{eq:ode_system}
\end{equation}
with initial conditions $A_{\mathrm{dep},i}(0) = d_i$ and $A_{\mathrm{cen},i}(0)=0$.
Here $k_{a,i}$ (h$^{-1}$) is the first-order absorption rate, $\mathrm{CL}_i$
(L/h) is the apparent clearance, and $V_i$ (L) is the apparent volume of
distribution. All three are assumed to follow a log-normal random effect structure
to enforce their positivity:
\begin{equation}
\begin{aligned}
k_{a,i} &\sim \mathrm{LogNormal}(\mu_{k_a},\sigma_{k_a}^2), \\
\mathrm{CL}_i &\sim \mathrm{LogNormal}\!\left(
\mu_{\mathrm{CL}} + 0.75\log\!\left(\frac{w_i}{70}\right),
\sigma_{\mathrm{CL}}^2
\right), \\
V_i &\sim \mathrm{LogNormal}(\mu_V,\sigma_V^2),
\end{aligned}
\label{eq:re_dist}
\end{equation}
where $w_i$ is the body weight of subject $i$ (kg) and 70\,kg is chosen as the reference
weight. The allometric exponent of~0.75 follows established pharmacometric
practice \citep{wang2012bodyweight}. Observed plasma concentrations $C_{ij}$ at
time $t_{ij}$ are linked to the \gls{ode} solution by
\begin{equation}
  C_{ij} \mid k_{a,i},\mathrm{CL}_i,V_i
    \;\sim\; \mathcal{N}\!\left(
      \frac{A_{\mathrm{cen},i}(t_{ij})}{V_i},\;\sigma_C^2
    \right),
  \label{eq:obs_C}
\end{equation}
The depot and central compartments of Equation~\ref{eq:ode_system} form a linear
system whose central-compartment amount has the closed-form Bateman solution
\begin{equation}
  A_{\mathrm{cen},i}(t) = d_i\,\frac{k_{a,i}}{k_{a,i} - k_{e,i}}
    \left(e^{-k_{e,i}\,t} - e^{-k_{a,i}\,t}\right),
  \qquad k_{e,i} = \frac{\mathrm{CL}_i}{V_i},
  \label{eq:bateman}
\end{equation}
which we evaluate directly in place of numerical integration. This yields seven
estimable parameters: the population log-means $\mu_{k_a}$, $\mu_{\mathrm{CL}}$,
$\mu_{V}$, the random-effect standard deviations $\sigma_{k_a}$,
$\sigma_{\mathrm{CL}}$, $\sigma_{V}$, and the residual error $\sigma_C$. Each
population parameter additionally carries a uniform prior, which bounds it for
frequentist estimation and serves as its prior for Bayesian estimation, so a
single specification supports both. The \code{@Model} macro encodes the
closed-form solution (Equation~\ref{eq:bateman}), the random-effect structure
(Equation~\ref{eq:re_dist}), and the observation model (Equation~\ref{eq:obs_C})
through a helper function:

\begin{CodeInput}
model = @Model begin
    @helpers begin
        bateman_center(t, dose, ka, ke) = abs(ka - ke) < 1e-8 ?
            dose * ka * t * exp(-ka * t) :
            dose * ka / (ka - ke) * (exp(-ke * t) - exp(-ka * t))
    end
    @fixedEffects begin
        ka_mean       = RealNumber(0.0,  prior = Uniform(-2, 2))
        CL_mean       = RealNumber(-0.5, prior = Uniform(-3, 3))
        V_mean        = RealNumber(4.0,  prior = Uniform(1, 4))
        sigma_ka      = RealNumber(2.0, scale = :log, prior = Uniform(0, 4))
        sigma_CL      = RealNumber(2.0, scale = :log, prior = Uniform(0, 4))
        sigma_V       = RealNumber(2.0, scale = :log, prior = Uniform(0, 4))
        sigma_C_error = RealNumber(2.0, scale = :log, prior = Uniform(0, 4))
    end
    @covariates begin
        t  = Covariate()
        d  = ConstantCovariate(; constant_on = :id)
        wt = ConstantCovariate(; constant_on = :id)
    end
    @randomEffects begin
        ka     = RandomEffect(LogNormal(ka_mean, sigma_ka); column = :id)
        CL     = RandomEffect(
                     LogNormal(CL_mean + 0.75 * log(wt / 70.0), sigma_CL);
                     column = :id)
        V      = RandomEffect(LogNormal(V_mean, sigma_V); column = :id)
    end
    @formulas begin
        C ~ Normal(bateman_center(t, d, ka, CL / V) / V, sigma_C_error)
    end
end

dm = DataModel(model, df; primary_id = :id, time_col = :t)
\end{CodeInput}

The \code{bateman\_center} helper implements the closed-form solution of
Equation~\ref{eq:bateman}. The four standard deviations are estimated on the log
scale (\code{scale = :log}), which enforces positivity by construction.
After the model definition, \code{DataModel(model, df; primary\_id = :id, time\_col = :t)} attaches the model to the
data.

\subsubsection{Estimation and uncertainty quantification}

Parameter estimation uses \gls{mcem} with \code{store\_diagnostics = true} to record
per-iteration parameter trajectories and Wald-type 95\% confidence
intervals are obtained from the observed Fisher information
matrix at the \gls{mcem} estimate:

\begin{CodeInput}
Random.seed!(12243)
fit = fit_model(dm, MCEM(maxiters = 60, store_diagnostics = true))
uq = compute_uq(fit)
summarize(uq)
\end{CodeInput}

\begin{CodeOutput}
UQResultSummary
=======================================================================
Overview
  backend                             : wald
  source_method                       : mcem
  inference                           : frequentist
  scale                               : natural
  interval level                      : 0.95
  parameters shown (reported / total) : 7 / 7

Parameter uncertainty summary
  parameter          Estimate    Std. Error      CI Lower      CI Upper
  ----------------------------------------------------------------------
  ka_mean             -0.5097        0.1943       -0.8859       -0.1144
  CL_mean             -2.0484         0.038       -2.1259       -1.9761
  V_mean               2.0526        0.0475        1.9585        2.1456
  sigma_ka             0.6445        0.0269        0.5933        0.7009
  sigma_CL             0.1706        0.0342        0.1163        0.2495
  sigma_V              0.2205        0.0382        0.1583        0.3021
  sigma_C_error        1.0499        0.0577        0.9403        1.1731
\end{CodeOutput}

The population estimates correspond to a median absorption rate of
approximately $0.60$~h$^{-1}$, a median clearance of $0.13$~L/h at the
70~kg reference weight, and a median volume of distribution of
$7.8$~L. The confidence intervals for clearance and volume are relatively
narrow, indicating good identifiability, whereas the wider interval for
the absorption rate reflects the limited information available during the
early absorption phase. These are frequentist point estimates with Wald-type
intervals. Because the estimator is selected by a single argument to
\code{fit\_model}, the same model is also fit by \gls{mcmc} to obtain a full
Bayesian posterior, which we examine in Section~\ref{sec:warfarin_bayes}.

\subsubsection{Results and diagnostics}
\begin{figure}[t!]
  \centering
  \includegraphics[width = \textwidth]{%
    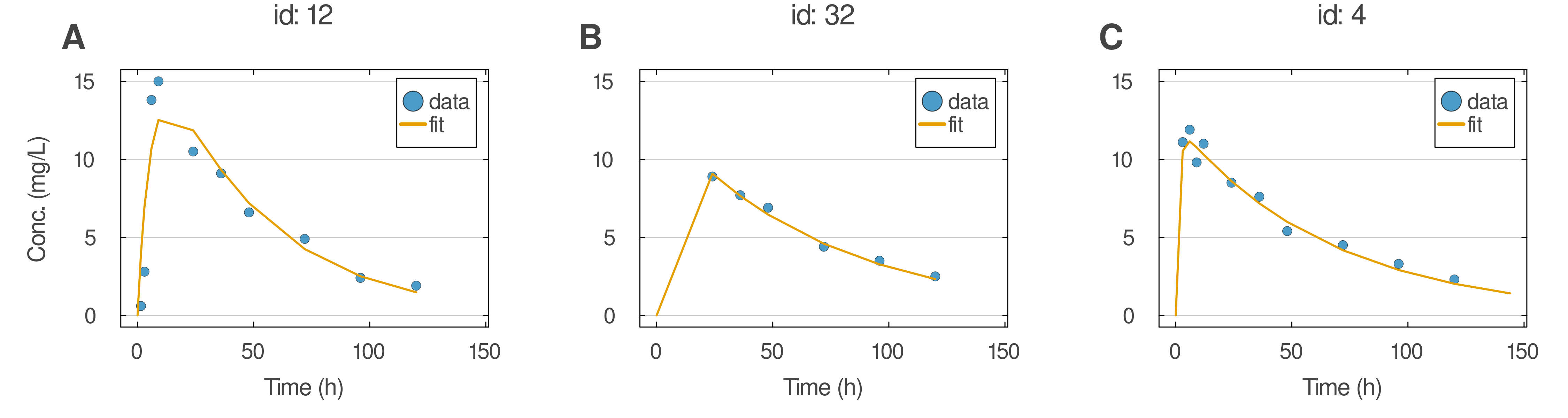}\\[4pt]
  \includegraphics[width = \textwidth]{%
    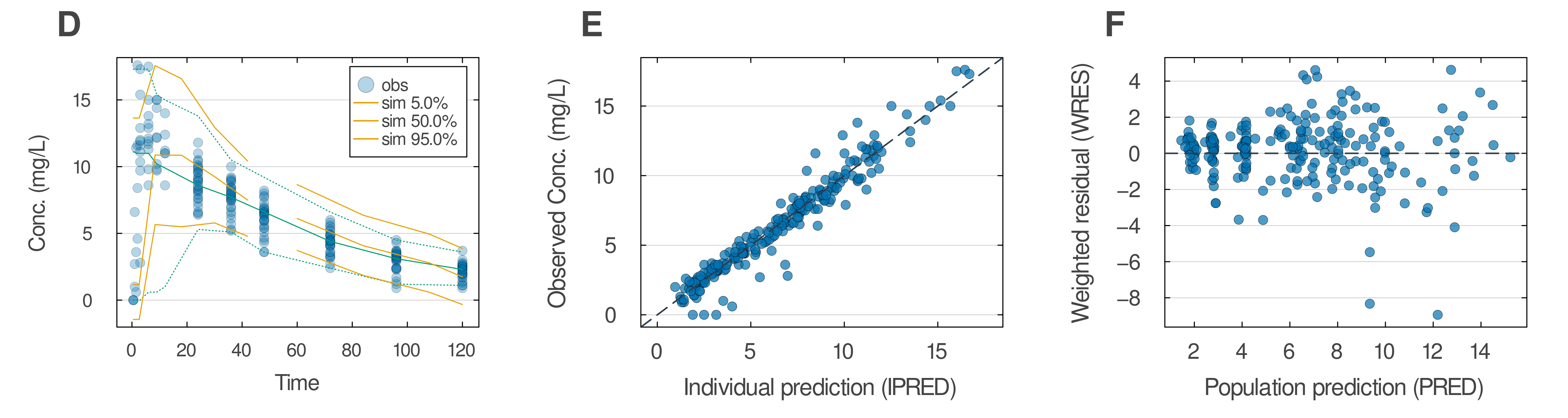}
  \caption{Model fit diagnostics for the warfarin analysis. (A)--(C)~Posterior predictions (curves) overlaid on observed plasma
    concentrations (dots) for three randomly selected subjects (top row), produced with \code{plot\_fits}.
    (D)~Visual predictive check with 5th, 50th, and 95th simulated
    percentile envelopes (shaded bands) and observed sample quantiles (solid
    lines) based on 1000 simulations, produced with \code{plot\_vpc}. (E)~Observed concentration against \gls{ipred} evaluated at
    \glspl{ebe}, produced with \code{plot\_dv\_ipred}. (F)~\Gls{wres}
    against \gls{pred}, produced with \code{plot\_wres\_pred}.}
  \label{fig:warfarin_diagnostics}
\end{figure}

Individual and population calibration are assessed jointly in
Figure~\ref{fig:warfarin_diagnostics}.
The individual fit is examined first, for three randomly selected subjects:

\begin{CodeInput}
fig3_idx = sort(randperm(MersenneTwister(42), 32)[1:3])
fig3 = plot_fits(fit; individuals_idx = fig3_idx, ncols = 3, style = my_style,
    kwargs_subplot = (xlabel = "Time (h)", ylabel = "Concentration (mg/L)"))
\end{CodeInput}

Population-level calibration is then assessed with a \gls{vpc} and two goodness-of-fit
comparisons of predictions, residuals, and observed values:

\begin{CodeInput}
p_vpc   = plot_vpc(fit; n_simulations = 1000, percentiles = [5, 50, 95],
                   style = my_style)
p_ipred = plot_dv_ipred(fit; style = my_style)
p_wres  = plot_wres_pred(fit; style = my_style)
\end{CodeInput}

The fitted model captures the absorption peak and the subsequent log-linear elimination
for all three subjects (Figure~\ref{fig:warfarin_diagnostics}A--C), and the between-subject
differences in peak height and elimination slope reflect the estimated inter-individual
variability in \code{ka}, \code{CL}, and \code{V}. The \gls{vpc}
(Figure~\ref{fig:warfarin_diagnostics}D) confirms that the population model reproduces the
marginal variability of the data across the full time course. Predictions evaluated at the
\glspl{ebe} agree closely with the observed concentrations
(Figure~\ref{fig:warfarin_diagnostics}E). The \glspl{wres} are approximately symmetric about zero with no systematic trend
against the \gls{pred} (Figure~\ref{fig:warfarin_diagnostics}F), indicating that the
additive residual error model is appropriate for this dataset.

The \gls{mcem} convergence trajectories and the inter-individual variability of the three random
effects, including the empirical-Bayes shrinkage and the body-weight dependence of clearance, are
reported in Appendix~\ref{app:warfarin_extra}.

\subsubsection{Bayesian analysis}
\label{sec:warfarin_bayes}

The model defined above already declares a uniform prior on every population
parameter, so a full Bayesian analysis needs no change to the specification, only
a different estimator. We draw from the posterior with the
\gls{nuts}~\citep{hoffman2014no} through the \pkg{Turing.jl} backend
\citep{turing, turingjl}, and the closed-form solution
(Equation~\ref{eq:bateman}) keeps the sampling inexpensive:

\begin{CodeInput}
fit_mcmc = fit_model(dm, MCMC(); rng = MersenneTwister(3))
\end{CodeInput}

\begin{figure}[t!]
  \centering
  \includegraphics[width = \textwidth]{%
    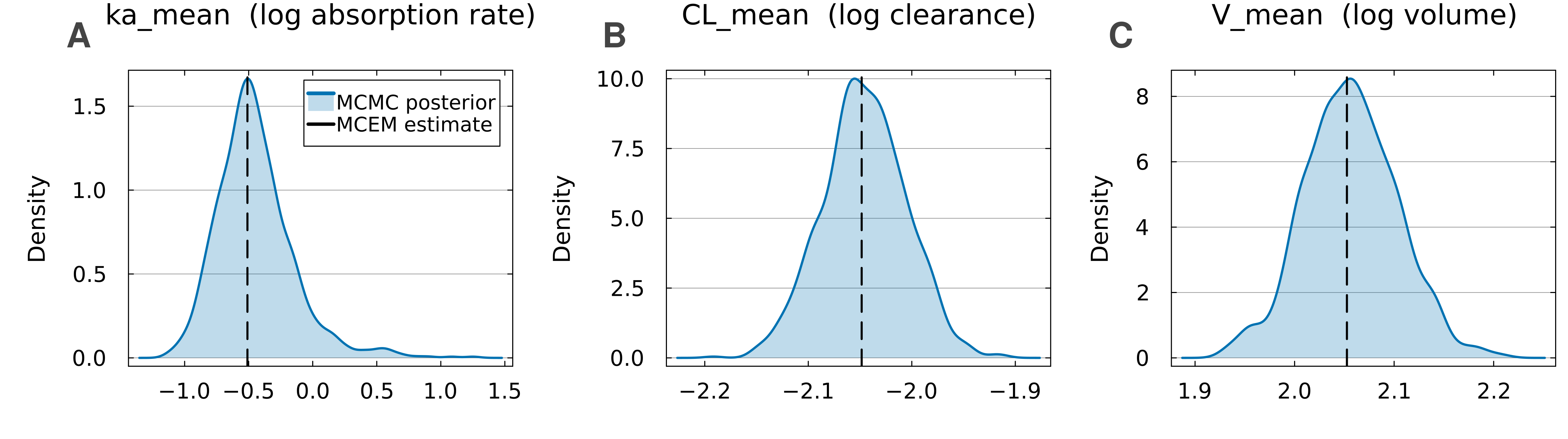}
  \caption{Marginal posterior distributions of the population log-means from the
    Bayesian fit (\gls{mcmc} with the \gls{nuts}), with the corresponding
    \gls{mcem} point estimates as dashed lines: (A)~$\mu_{k_a}$,
    (B)~$\mu_{\mathrm{CL}}$, (C)~$\mu_V$.}
  \label{fig:warfarin_mcmc}
\end{figure}

Figure~\ref{fig:warfarin_mcmc} compares the marginal posteriors of the
population log-means with the corresponding \gls{mcem} estimates. The
posterior modes closely match the \gls{mcem} point estimates, indicating good
agreement between the frequentist and Bayesian analyses. Clearance and
volume are well identified, with narrow, approximately symmetric
posteriors. In contrast, the absorption-rate posterior is broader
and right-skewed, reflecting the
limited information available on the absorption phase. This asymmetry is
captured by the posterior distribution but not by a point estimate or the
symmetric Wald interval from the MCEM estimation.
Posterior summaries for the random-effect and outcome distribution
standard deviations are reported in
Appendix~\ref{app:warfarin_bayes}.

\subsection{Warfarin pharmacodynamics: a learnable concentration-effect
  function}
\label{sec:warfarin_pkpd}

The one-compartment model of Section~\ref{sec:warfarin} captures warfarin
plasma concentrations well, but the same dataset also contains \gls{inr}
measurements, which we have ignored in the previous model.
A standard \gls{pd} model links concentration to an indirect-response
turnover model through a saturable \code{Emax} effect
\citep{dayneka1993comparison}. Alternatively, this relationship can be
learned directly from the data.
\nl{} supports this by allowing users to embed differentiable machine-learning
components directly into the turnover model.
We demonstrate this with neural-network and soft-tree models and use
\gls{cv} to compare them against the parametric model, showing how a learned
effect function serves as a diagnostic for the assumed concentration-effect
shape even when predictive performance is unchanged.

\subsubsection{Learnable concentration-effect dynamics}

This \gls{pd} model extends the \gls{pk} model of Section~\ref{sec:warfarin}. The depot and central compartments of
Equation~\ref{eq:ode_system} are kept unchanged, and an indirect-response turnover
compartment is added on top, driven by the plasma concentration
$C_i(t) = A_{\mathrm{cen},i}(t)/V_i$ from that submodel. Writing $R_i$ for the
response of subject $i$, $R_{0,i}$ for its baseline, and $k_{\mathrm{out},i}$ for the
turnover rate, the drug modifies the loss of the response through
\begin{equation}
\frac{dR_i}{dt}
  = k_{\mathrm{out},i}\Bigl(R_{0,i} - R_i(t)\,\bigl(1 + \mathcal{E}(C_i(t))\bigr)\Bigr),
\qquad R_i(0) = R_{0,i},
\label{eq:pd_turnover}
\end{equation}
where $\mathcal{E}$ is a concentration-effect function. For this concentration-effect function, we compare a parametric saturable effect
with a learnable one,
\begin{equation}
\mathcal{E}(C) = \frac{E_{\max}\,C}{\mathrm{EC}_{50} + C}
\qquad\text{or}\qquad
\mathcal{E}(C) = C\,\zeta\!\bigl(\mathrm{ML}(C/c_0)\bigr),
\label{eq:effect}
\end{equation}
where $\mathrm{ML}$ is a neural network or a soft decision tree, $\zeta$ the softplus
function ensuring a nonnegative effect, $c_0$ a fixed concentration scale, and the
leading factor $C$ ensures no effect at zero concentration. For conciseness in the main manuscript, the full
\code{@Model} specification of the parametric \code{Emax} model is given in
Appendix~\ref{app:warfarin_pkpd_param}. The neural network-based model embeds a small
\pkg{SimpleChains.jl} network in the response dynamics:

\begin{CodeInput}
chainE = SimpleChain(static(1), TurboDense(tanh, 4), TurboDense(identity, 1))

model_nn = @Model begin
    @helpers begin
        softplus(u) = u > 20 ? u : log1p(exp(u))
    end
    @fixedEffects begin
        ka_mean    = RealNumber(0.0); CL_mean    = RealNumber(-2.0)
        V_mean     = RealNumber(2.0); kout_mean  = RealNumber(-3.0)
        sigma_ka   = RealNumber(0.3, scale = :log)
        sigma_CL   = RealNumber(0.3, scale = :log)
        sigma_V    = RealNumber(0.3, scale = :log)
        sigma_C    = RealNumber(1.0, scale = :log)
        sigma_kout = RealNumber(0.3, scale = :log)
        sigma_R    = RealNumber(5.0, scale = :log)
        ml_params  = NNParameters(chainE; function_name = :ML)
    end
    @covariates begin
        t  = Covariate()
        d  = ConstantCovariate(; constant_on = :id)
        wt = ConstantCovariate(; constant_on = :id)
        R0 = ConstantCovariate(; constant_on = :id)
    end
    @randomEffects begin
        ka   = RandomEffect(LogNormal(ka_mean, sigma_ka); column = :id)
        CL   = RandomEffect(
                   LogNormal(CL_mean + 0.75 * log(wt / 70.0), sigma_CL);
                   column = :id)
        V    = RandomEffect(LogNormal(V_mean, sigma_V); column = :id)
        kout = RandomEffect(LogNormal(kout_mean, sigma_kout); column = :id)
    end
    @DifferentialEquation begin
        eff(t) = (center / V) *
                 softplus(ML([(center / V) / 5.0], ml_params)[1])
        D(depot)    ~ -ka * depot
        D(center)   ~  ka * depot - (CL / V) * center
        D(response) ~  kout * (R0 - response * (1 + eff(t)))
    end
    @initialDE begin
        depot = d; center = 0.0; response = R0
    end
    @formulas begin
        C ~ Normal(center(t) / V, sigma_C)
        R ~ Normal(response(t), sigma_R)
    end
end
\end{CodeInput}

The soft-tree model has the same structure and differs only in the effect
component. Because the \code{@Model} macro can extend an existing model, the
soft-tree model is obtained from the neural model by overriding a single fixed
effect:

\begin{CodeInput}
model_st = @Model model_nn begin
    @fixedEffects begin
        ml_params = SoftTreeParameters(1, 2; function_name = :ML)
    end
end
\end{CodeInput}

Reusing the function name \code{:ML} leaves the differential-equation, observation,
and random-effect blocks of \code{model\_nn} unchanged, so swapping the network for
a depth-two soft decision tree is a few-line change that leaves all other model
components unchanged.

\subsubsection{Cross-validated comparison}

In the following example, we compare the three models using 10-fold
leave-subjects-out \gls{cv}. In each fold, entire subjects are held out
and their responses are predicted from the population distribution,
mimicking prediction for a new patient. Performance is evaluated by the
mean squared error of the held-out \gls{inr} response via the
\code{loss} argument. We score \gls{inr} rather than concentration
because the \gls{pk} submodel is identical across all models, so only
the \gls{pd} component differs. Setting \code{store\_results = true}
retains the fitted models from each fold, allowing the learned effect
function, predicted response, and their cross-validation uncertainty to
be reconstructed. The soft-tree model \code{model\_st} and the parametric
\code{Emax} reference of Appendix~\ref{app:warfarin_pkpd_param} are bound to the
same data and cross-validated with identical calls, so the three results are
directly comparable:

\begin{CodeInput}
dm_nn = DataModel(model_nn, df; primary_id = :id, time_col = :t)
mse_loss = (dist, y) -> (mean(dist) - y)^2

cv = cross_validate(dm_nn, 10; kind = :id, rng = MersenneTwister(1))
fit = fit_cv(cv, SAEM(maxiters = 200);
    unseen_re_mode = :mean, loss = mse_loss, store_results = true,
    rng = MersenneTwister(2))
\end{CodeInput}

\begin{figure}[t!]
  \centering
  \includegraphics[width = \textwidth]{%
    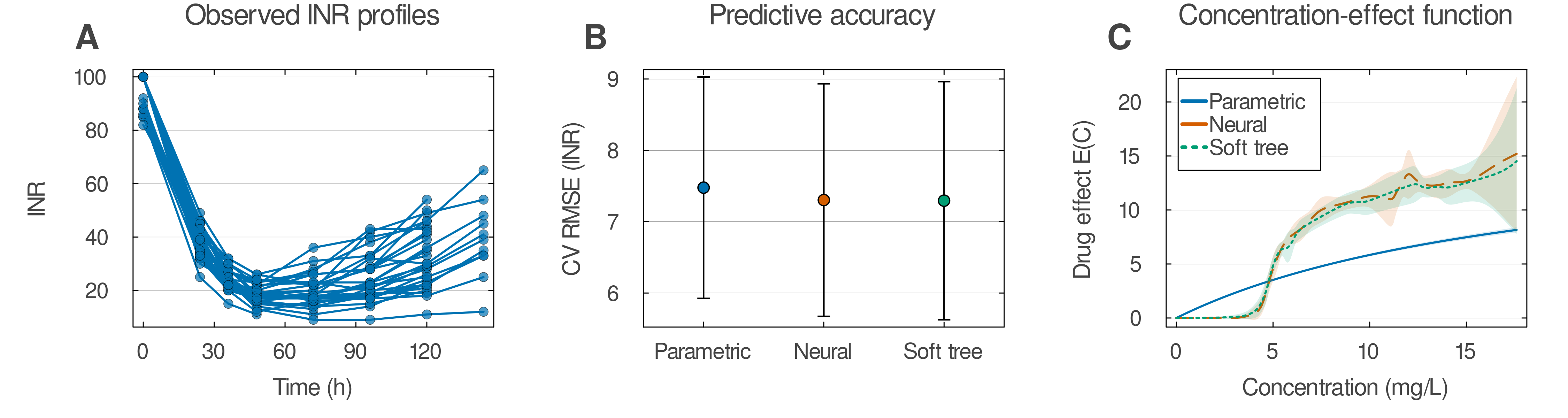}
  \caption{Warfarin anticoagulant response and pharmacodynamic model comparison.
    (A)~Observed \gls{inr} responses over time for all subjects. (B)~Cross-validated
    \gls{rmse} on the held-out \gls{inr} response for each model under 10-fold
    leave-subjects-out cross-validation (mean and 95\% confidence interval across
    folds). (C)~Estimated concentration-effect function $\mathcal{E}(C)$ for the parametric
    \code{Emax} model and the learned neural-network and soft-tree effects, with
    pointwise 95\% confidence bands across folds.}
  \label{fig:warfarin_pkpd}
\end{figure}

The \gls{inr} response declines after dosing and recovers as warfarin
clears, with greater between-subject variability during recovery
(Figure~\ref{fig:warfarin_pkpd}A). The estimated concentration-effect
relationships carry the main finding of the analysis
(Figure~\ref{fig:warfarin_pkpd}C). The neural network and soft tree, fitted
independently, recover similar threshold-shaped effects that remain close to
zero below roughly 4~mg/L before increasing steeply, a qualitatively different
shape from the gradual response implied by the parametric \code{Emax} model.
This learned shape is the diagnostic payoff of the flexible components, because
the agreement of two independent function approximators on a threshold form is
stronger evidence for that form than either model alone, and it identifies a
plausible alternative to the saturable \code{Emax} assumption that a purely
parametric workflow would not surface. The three models nonetheless achieve
nearly identical held-out performance, with overlapping cross-validated
\gls{rmse} intervals (Figure~\ref{fig:warfarin_pkpd}B), because the effect shape
trades off against the turnover rate $k_{\mathrm{out}}$. The \code{Emax} model
compensates for its gradual effect with a turnover rate of
$0.0086~\mathrm{h}^{-1}$ [0.0084, 0.0088], approximately twice that of the
neural network ($0.0040$ [0.0036, 0.0043]) and soft tree ($0.0041$ [0.0037,
0.0046]), so all three generate nearly identical \gls{inr} predictions despite
markedly different concentration-effect curves. The flexible models therefore do
not improve prediction on this dataset, but they earn their place as a
diagnostic that reveals structure in the concentration-effect relationship, a
conclusion that remains suggestive rather than definitive because the flexible
models do not share the \code{Emax} parameterization.

\subsection[Fish growth: normalizing flows as flexible random-effects
  distributions]{Fish growth: normalizing flows as flexible
  random-effects distributions}
\label{sec:flow}

The previous two examples relied on Gaussian random effects. Partial migration can instead produce multimodal growth-rate distributions that violate this assumption \citep{hegemann2019physiological}.
Normalizing flows address this by learning flexible, potentially
multimodal random-effects distributions from the data.
In \nl{}, a normalizing flow can directly replace a Gaussian
random-effects distribution without changing the rest of the model.
We compare Gaussian and flow-based random-effects models on simulated data
using \gls{saem}.

\subsubsection{Data}

The dataset of \citet{gillanders2015partial} comprises individual growth records for
167 black bream (\textit{Acanthopagrus butcheri}), a long-lived estuarine sparid, sampled
in the Coorong Estuary, South Australia, between 2008 and 2012.
Captured fish span ages 5 to 32 years and lengths of 283 to 470~mm, and the original
study classified each fish as resident (104) or migratory (63) from otolith microchemistry,
a partial-migration structure that motivates a multimodal growth-rate distribution. Rather
than fitting these records directly, we use them to calibrate a realistic simulation in
which the true random-effects distribution is known, so that the recovery of latent
sub-population structure can be assessed against a ground truth (Section~\ref{sec:fish_sim}).
The simulated morph labels are withheld from every estimation model and retained only to
color the figures, so any recovered bimodality must emerge from the fitted growth
trajectories alone.

\subsubsection{Model specification}

We model length at age with a \gls{vbgf}~\citep{vonBertalanffy1957quantitative}.
For fish $i = 1, \ldots, N$ observed at ages $t_{ij}$, $j = 1, \ldots, n_i$, length
$L_{ij}$ is described by
\begin{align}
  L_{ij} \mid L_{\infty,i},\, k_i
    &\;\sim\; \mathcal{N}\!\left(
        L_{\infty,i}\bigl(1 - e^{-k_i\, t_{ij}}\bigr),\;
        \sigma_y^2
      \right), \label{eq:vbgf_obs} \\[4pt]
  L_{\infty,i} &\;=\; L_{\infty,\mathrm{pop}}\,e^{b_{L,i}}, \qquad
    b_{L,i} \;\sim\; \mathcal{N}(0,\, \sigma_L^2), \label{eq:re_Linf} \\[4pt]
  k_i &\;=\; k_{\mathrm{pop}}\,e^{b_{k,i}}, \qquad
    b_{k,i} \;\sim\; \mathcal{D}, \label{eq:re_k}
\end{align}
where $L_{\infty,\mathrm{pop}} > 0$ is the population-level asymptotic length,
$k_{\mathrm{pop}} > 0$ is the population-level growth-rate coefficient, $\sigma_L > 0$
controls inter-individual variability in asymptotic length, and $\sigma_y > 0$ is the
residual standard deviation. 
The asymptotic-length random effect $b_{L,i}$ is Gaussian
throughout, whereas the growth-rate random effect $b_{k,i}$ is drawn from a distribution
$\mathcal{D}$ that we deliberately leave unspecified at this stage. This distribution is
the single component that varies across the remainder of the section. We first instantiate
$\mathcal{D}$ as a known bimodal mixture to generate simulated data
(Section~\ref{sec:fish_sim}), then estimate it from those data under two competing
assumptions, a Gaussian family and a flexible normalizing flow
(Section~\ref{sec:fish_fit}).

\subsubsection{Data simulation}
\label{sec:fish_sim}

To test whether a flexible random-effects distribution recovers latent structure when it
is genuinely present, we simulate from the model of
Equations~\ref{eq:vbgf_obs}--\ref{eq:re_k} with $\mathcal{D}$ set to a known bimodal
distribution
\begin{equation}
  b_{k,i} \;\sim\; \frac{1}{2}\,\mathcal{N}(-\mu,\,\sigma_c^2)
              \;+\; \frac{1}{2}\,\mathcal{N}(\mu,\,\sigma_c^2),
  \qquad \mu = 0.70,\quad \sigma_c = 0.20,
  \label{eq:true_mixture}
\end{equation}
representing a slow-growing morph centered at $-\mu$ and a fast-growing morph centered at $\mu$.

The data-generating parameters are calibrated to the real \citet{gillanders2015partial} data by a
Gaussian \gls{saem} fit, giving an asymptotic length near $705$~mm, growth coefficient near
$0.104$~year$^{-1}$, and residual standard deviation near $15$~mm.

In \pkg{NoLimits.jl} the mixture is realized
through a binary covariate \code{morph} that selects the component mean, so that over an
even split of morphs the marginal distribution of $b_{k,i}$ matches
Equation~\ref{eq:true_mixture}. The \code{morph} label is supplied only to the simulation
model and withheld from every estimation model, which observes only \code{fishid},
\code{age}, and \code{length}, so any recovered bimodality must emerge from the fitted
growth curves rather than from a known grouping. The simulation model encodes the mixture in its
\code{@randomEffects} block:

\begin{CodeInput}
model_sim = @Model begin
    @covariates begin
        age   = Covariate()
        morph = ConstantCovariate(; constant_on = :fishid)
    end
    @fixedEffects begin
        L_inf_pop  = RealNumber(705.0, scale = :log)
        k_pop      = RealNumber(0.104, scale = :log)
        sigma_L    = RealNumber(0.10,  scale = :log)
        sigma_y    = RealNumber(15.0,  scale = :log)
        mu_bim     = RealNumber(0.70)
        sigma_comp = RealNumber(0.20,  scale = :log)
    end
    @randomEffects begin
        b_k = RandomEffect(Normal((2 * morph - 1) * mu_bim, sigma_comp);
                             column = :fishid)
        b_L = RandomEffect(Normal(0.0, sigma_L); column = :fishid)
    end
    @formulas begin
        length ~ Normal(
            L_inf_pop * exp(b_L) *
                (1 - exp(- k_pop * exp(b_k) * age)),
            sigma_y)
    end
end
\end{CodeInput}

The component-mean term \code{(2 * morph - 1) * mu\_bim} evaluates to $-\mu$ for \code{morph = 0} and
$\mu$ for \code{morph = 1}, encoding the mixture of Equation~\ref{eq:true_mixture}. The model is
attached to a template \code{DataFrame} of 100 fish split evenly between the morphs and observed at
ages 2--15 years, constructed as shown in the replication materials. Observations are drawn with
\code{simulate\_data}, the \pkg{NoLimits.jl} function that samples the random effects from the prior
and appends their true values to the returned \code{DataFrame}:

\begin{CodeInput}
dm_sim = DataModel(model_sim, df_template; 
                   primary_id = :fishid, time_col = :age)
df_sim = simulate_data(dm_sim; rng = MersenneTwister(42))
\end{CodeInput}

The simulated length-at-age trajectories separate into two bundles, the fast morph reaching
markedly larger lengths than the slow morph, yet they overlap at the youngest ages where growth has
not yet diverged, so the morph of an individual fish is not perfectly identifiable from its
trajectory alone. The marginal density of the growth-rate random effect is the equally weighted
bimodal mixture of Equation~\ref{eq:true_mixture}, the latent structure that a Gaussian
random-effects distribution cannot represent and that the normalizing flow is designed to recover. It
appears as the dashed reference in Figure~\ref{fig:fish_recovery}A.

\subsubsection{Model fitting and comparison}
\label{sec:fish_fit}

We now instantiate $\mathcal{D}$ in two estimation models and fit both to the simulated
data with \gls{saem}. The Gaussian baseline places
\begin{equation}
  b_{k,i} \;\sim\; \mathcal{N}(0,\, \sigma_k^2), \label{eq:gauss_re}
\end{equation}
with $\sigma_k > 0$, while the flow model replaces
Equation~\ref{eq:gauss_re} with a normalizing planar flow $p_\psi$ parameterized by
weights $\psi$, whose density follows the change-of-variables form of Equation~\ref{eq:flow}. The complete parameter vector is
$\theta = (L_{\infty,\mathrm{pop}},\, k_{\mathrm{pop}},\, \sigma_L,\, \sigma_y,\, \sigma_k)$
for the Gaussian model and
$\theta = (L_{\infty,\mathrm{pop}},\, k_{\mathrm{pop}},\, \sigma_L,\, \sigma_y,\, \psi)$
for the flow model. The Gaussian specification translates directly into the \code{@Model}
macro:

\begin{CodeInput}
model_gauss = @Model begin
    @covariates begin; age = Covariate(); end
    @fixedEffects begin
        L_inf_pop = RealNumber(705.0, scale = :log)
        k_pop     = RealNumber(0.104, scale = :log)
        sigma_L   = RealNumber(0.10,  scale = :log)
        sigma_k   = RealNumber(0.40,  scale = :log)
        sigma_y   = RealNumber(15.0,  scale = :log)
    end
    @randomEffects begin
        b_k = RandomEffect(Normal(0.0, sigma_k); column = :fishid)
        b_L = RandomEffect(Normal(0.0, sigma_L); column = :fishid)
    end
    @formulas begin
        length ~ Normal(
            exp(log(L_inf_pop) + b_L) *
                (1 - exp(-exp(log(k_pop) + b_k) * age)),
            sigma_y)
    end
end
\end{CodeInput}

The normalizing flow specification makes three targeted changes to the Gaussian model:
\code{sigma\_k} is replaced by \code{NPFParameter(1, 4; seed = 42, calculate\_se = false)},
which parameterizes the weights of four successive planar transformation layers, the
\code{b\_k} random effect distribution is specified as \code{NormalizingPlanarFlow(psi)} in place of a
Gaussian distribution, and the random effect is now a vector, so it is indexed with square brackets
as \code{b\_k[1]} in the formula. All remaining declarations match \code{model\_gauss}, so these three
changes define \code{model\_flow}. Both specifications are attached to the simulated data via
\code{DataModel}:

\begin{CodeInput}
dm_gauss = DataModel(model_gauss, df_sim;
                     primary_id = :fishid, time_col = :age)
dm_flow  = DataModel(model_flow,  df_sim; 
                     primary_id = :fishid, time_col = :age)
\end{CodeInput}

Both models are fit with \gls{saem} for 800 iterations. The flow's \code{k\_pop} is held at the Gaussian estimate through the \code{constants} keyword, because the flow random effect, unlike the Gaussian one, is not constrained to be centered at zero. A location shift can therefore move freely between \code{k\_pop} and the random-effect distribution, leaving the two not jointly identifiable. The fitted fixed effects are then placed side by side with \code{compare\_parameters}, which aligns the parameters shared by the two models and marks with a dash any parameter that one model does not have:

\begin{CodeInput}
Random.seed!(42)
fit_gauss = fit_model(dm_gauss, SAEM(maxiters = 800))
fit_flow  = fit_model(dm_flow,  SAEM(maxiters = 800);
                      constants = (k_pop = get_params(fit_gauss).k_pop,))
compare_parameters(fit_gauss, fit_flow; labels = ["Gaussian", "flow"])
\end{CodeInput}

\begin{CodeOutput}
ParameterComparison
===============================
  parameter  Gaussian      flow
-------------------------------
  L_inf_pop  718.4175  703.1301
  k_pop        0.1014    0.1014
  sigma_L      0.1083    0.1208
  sigma_k      0.7272         -
  sigma_y        15.4   15.3597
\end{CodeOutput}

\begin{figure}[t!]
  \centering
  \includegraphics[width = \textwidth]{%
    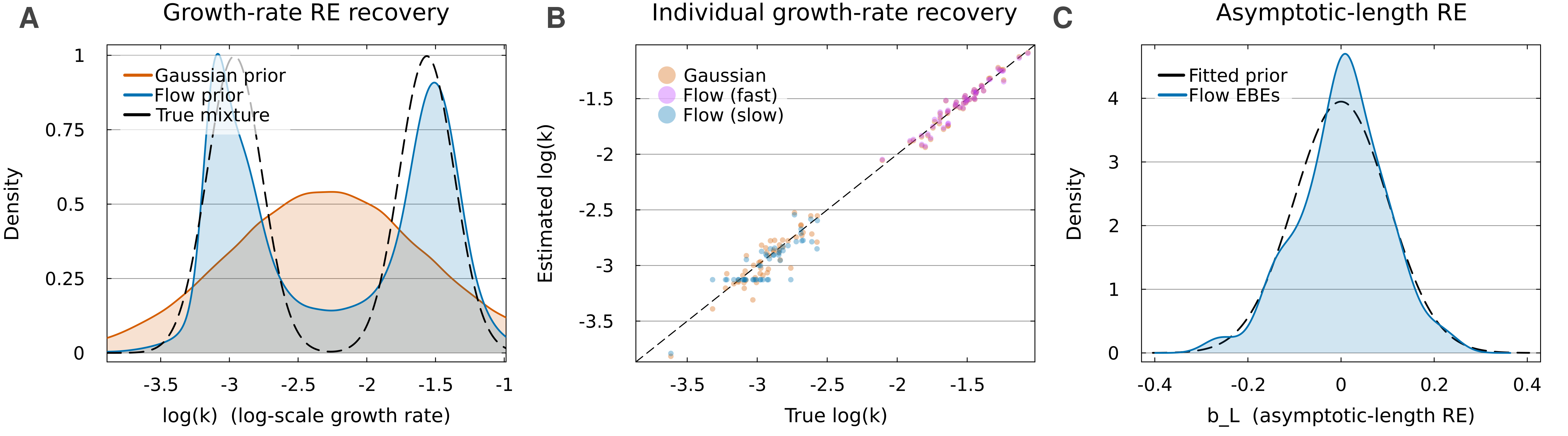}
  \caption{Random-effect distribution recovery on simulated data.
    (A)~Fitted growth-rate random-effect priors on the $\log(k)$ scale: the
    true generating mixture (black dashed), the fitted Gaussian prior (red), and
    the fitted normalizing-flow prior (blue filled). Both fitted priors are shown
    as kernel density estimates of draws from the corresponding random-effect
    distribution.
    (B)~Estimated versus true $\log(k)$ for both models, with flow-model
    \glspl{ebe} colored by true morph (orange: fast, blue: slow), Gaussian-model
    \glspl{ebe} in red, and the identity line dashed.
    (C)~Flow-model \glspl{ebe} for the asymptotic-length random effect
    $b_L$ (blue) overlaid on the fitted Gaussian prior (black dashed).}
  \label{fig:fish_recovery}
\end{figure}

The two models recover markedly different random-effect priors
(Figure~\ref{fig:fish_recovery}A). The Gaussian prior is unimodal and
places substantial density between the two true modes, whereas the flow
recovers the underlying bimodal structure. Both models estimate
individual growth-rate \glspl{ebe} with similar accuracy
(Figure~\ref{fig:fish_recovery}B), indicating that the data are
sufficiently informative at the individual level. The main advantage of
the flow therefore lies in its population-level representation: samples
from the flow reproduce both subpopulations, while samples from the
Gaussian prior concentrate unrealistically between them. In contrast, the
asymptotic-length random effect $b_L$, modeled as Gaussian in both
cases, is recovered consistently (Figure~\ref{fig:fish_recovery}C),
demonstrating that Gaussian and flow-based random effects can be combined
within the same model.
\begin{CodeInput}
plot_random_effects_pdf(fit_gauss; re_names = :b_k, style = my_style)
plot_random_effects_pdf(fit_flow;  re_names = :b_k, style = my_style)
\end{CodeInput}

\section{Comparison to other software packages}
\label{sec:comparison}
This section positions \pkg{NoLimits.jl} within the open-source ecosystem
for \gls{nlme} and longitudinal modeling.

\subsection{Scope and the open-source landscape}

We restrict the comparison to open-source software. Commercial tools such
as \pkg{NONMEM} \citep{bauer2019nonmem}, \pkg{Monolix}
\citep{Monolix2024R1}, and \pkg{Pumas}
\citep{rackauckas2020accelerated} are therefore excluded. Among these,
\pkg{Pumas} offers a comparably broad range of model classes, random-effects distributions, and inference paradigms. Thus, the contribution of \nl{} is best understood as making a similar breadth of capabilities available in an open-source framework, rather than as introducing all of these capabilities for the first time.
We also do not compare directly against general-purpose probabilistic
programming languages such as \pkg{Stan} \citep{stan},
\pkg{Turing.jl} \citep{turing, turingjl}, and \pkg{PyMC}
\citep{pymc2023}. These provide flexible inference engines rather than
domain-specific \gls{nlme} frameworks. In particular, \pkg{NoLimits.jl}
builds on \pkg{Turing.jl} for Bayesian inference, generating
\pkg{Turing.jl} models at runtime and reusing its sampling algorithms.
Rather than competing with these tools, \pkg{NoLimits.jl} provides a
higher-level modeling framework that integrates mixed-effects, \gls{ode},
and Markov models together with specialized estimation and diagnostic
methods.

The most closely related open-source packages are the \proglang{R}
packages \pkg{nlmixr2} \citep{fidler2025nlmixr2} and \pkg{saemix}
\citep{comets2017parameter}, which focus on pharmacometric mixed-effects
modeling, as well as specialized latent-state packages such as
\pkg{mHMMbayes} \citep{mHMMbayes}. The remainder of this section compares
\pkg{NoLimits.jl} primarily with \pkg{nlmixr2} and \pkg{saemix}. As
summarized in Table~\ref{tab:comparison}, \pkg{NoLimits.jl} unifies
a broader range of model classes, random-effects distributions, learned
components, and inference paradigms within a single composable framework.

\subsection[Close comparison to open-source tools]
{Close comparison: \pkg{nlmixr2} and \pkg{saemix}}

\begin{table}[htbp]
\centering
\footnotesize
\setlength{\tabcolsep}{5pt}
\renewcommand{\arraystretch}{1.25}
\begin{minipage}[t]{0.49\textwidth}
\begin{tabularx}{\linewidth}{@{}X>{\centering\arraybackslash}p{0.7cm}>{\centering\arraybackslash}p{0.7cm}>{\centering\arraybackslash}p{0.7cm}@{}}
\toprule
Feature & \rotatebox{90}{\pkg{NoLimits.jl}} & \rotatebox{90}{\pkg{nlmixr2}} & \rotatebox{90}{\pkg{saemix}} \\
\midrule
\multicolumn{4}{@{}l}{\textit{Model classes}} \\
\acrshort{ode}-based models & $\checkmark$ & $\checkmark$ & ($\checkmark$)\textsuperscript{a} \\
Closed-form models & $\checkmark$ & $\checkmark$ & $\checkmark$ \\
Markov models & $\checkmark$ & -- & -- \\
Time-to-event, censored & $\checkmark$ & $\checkmark$ & $\checkmark$ \\
\addlinespace[0.7em]
\multicolumn{4}{@{}l}{\textit{Random effects}} \\
Gaussian & $\checkmark$ & $\checkmark$ & $\checkmark$ \\
Skewed  & $\checkmark$ & $\checkmark$\textsuperscript{b} & $\checkmark$\textsuperscript{c} \\
Heavy-tailed ($t$, Laplace) & $\checkmark$ & -- & -- \\
Normalizing-flow & $\checkmark$ & -- & -- \\
\addlinespace[0.7em]
\multicolumn{4}{@{}l}{\textit{Observation models}} \\
Non-Gaussian observations & $\checkmark$ & $\checkmark$ & $\checkmark$
\textsuperscript{d} \\
\bottomrule
\end{tabularx}
\end{minipage}
\hfill
\begin{minipage}[t]{0.49\textwidth}
\begin{tabularx}{\linewidth}{@{}X>{\centering\arraybackslash}p{0.7cm}>{\centering\arraybackslash}p{0.7cm}>{\centering\arraybackslash}p{0.7cm}@{}}
\toprule
Feature & \rotatebox{90}{\pkg{NoLimits.jl}} & \rotatebox{90}{\pkg{nlmixr2}} & \rotatebox{90}{\pkg{saemix}} \\
\midrule
\multicolumn{4}{@{}l}{\textit{Machine Learning}} \\
Neural Networks & $\checkmark$ & ($\checkmark$)\textsuperscript{e} & -- \\
Soft Trees & $\checkmark$ & -- & -- \\
\addlinespace[0.7em]
\multicolumn{4}{@{}l}{\textit{Inference and uncertainty}} \\
Likelihood / \acrshort{em} & $\checkmark$ & $\checkmark$ & $\checkmark$ \\
Bayesian (\gls{mcmc}) & $\checkmark$ & -- & -- \\
Bootstrap uncertainty & -- & $\checkmark$ & $\checkmark$ \\
Automated covariate search & -- & $\checkmark$ & -- \\
\addlinespace[0.7em]
\multicolumn{4}{@{}l}{\textit{Platform and ecosystem}} \\
Language & \proglang{Julia} & \proglang{R} & \proglang{R} \\
Open-source & $\checkmark$ & $\checkmark$ & $\checkmark$ \\
NONMEM/Monolix import & -- & $\checkmark$ & -- \\
\bottomrule
\end{tabularx}
\end{minipage}
\caption{Feature comparison of \pkg{NoLimits.jl}, \pkg{nlmixr2}, and
\pkg{saemix}. Check marks denote native support, bracketed check marks
conditional or extension-based support, and dashes no readily available support in the standard workflow.
\textsuperscript{a}~External \gls{ode} solver.
\textsuperscript{b}~Transformation-based support.
\textsuperscript{c}~Via \code{transform.par}.
\textsuperscript{d}~Via user-defined likelihoods
\citep{comets2026extending}.
\textsuperscript{e}~Via \pkg{pmxNODE} \citep{pmxnode}.
Versions: \pkg{NoLimits.jl}~0.1.0, \pkg{nlmixr2}~4.0.1,
\pkg{saemix}~3.5.}
\label{tab:comparison}
\end{table}

\pkg{nlmixr2} \citep{fidler2025nlmixr2} and \pkg{saemix}
\citep{comets2017parameter} are mature and widely used pharmacometric
\proglang{R} packages centered on \gls{saem} and, for \pkg{nlmixr2},
also \gls{focei}. The differences discussed below are therefore primarily
differences in scope: \pkg{nlmixr2} and \pkg{saemix} focus on
pharmacometric workflows, whereas \pkg{NoLimits.jl} emphasizes breadth
across model classes, random-effects distributions, learned components,
and inference paradigms for general and neural \gls{nlme}.

Table~\ref{tab:comparison} summarizes the main differences. The most
pronounced distinction lies in the random-effects distribution.
\pkg{nlmixr2} and \pkg{saemix} support Gaussian and transformed
distributions such as the log-normal, but not symmetric heavy-tailed or
flow-based random effects. \pkg{NoLimits.jl} additionally supports
Student's $t$, Laplace, and normalizing-flow distributions, enabling
flexible and multimodal random-effects models as illustrated in
Section~\ref{sec:flow}.

The supported model classes also differ. Neither \pkg{nlmixr2} nor
\pkg{saemix} provides native support for Markov models, whereas
\pkg{NoLimits.jl} includes discrete- and continuous-time Markov
observation models with hidden or observed states within the same
framework.

Neural-network integration is another point of departure. Through the
third-party \pkg{pmxNODE} package \citep{pmxnode}, \pkg{nlmixr2}
supports a restricted class of neural-\gls{ode} models, while
\pkg{saemix} offers no direct neural-network support. In contrast,
\pkg{NoLimits.jl} integrates arbitrary \pkg{Lux.jl} and \pkg{SimpleChains.jl} networks, as well as differentiable soft trees as
composable model components that can appear in structural, observation,
and random-effects models.

Finally, \pkg{NoLimits.jl} unifies likelihood-based, \gls{em}-based, and
Bayesian inference behind a common interface, including \gls{mcmc} through its \pkg{Turing.jl} backend.
Neither \pkg{nlmixr2} nor \pkg{saemix} provides Bayesian \gls{mcmc}
inference.

These advantages in flexibility are balanced by strengths of the
incumbents. \pkg{nlmixr2} and \pkg{saemix} benefit from extensive
validation, large user communities, and pharmacometric workflow tooling
such as bootstrap uncertainty quantification, automated covariate search,
and interoperability with established tools. For workflows that depend on
this ecosystem, they remain compelling choices.

\subsection{Decision guidance}

In summary, the comparison above points to complementary strengths.
\pkg{nlmixr2} and \pkg{saemix}
\citep{fidler2025nlmixr2, comets2017parameter} remain natural choices
for established pharmacometric workflows that rely on their mature
ecosystems and extensive validation, whereas \pkg{NoLimits.jl} targets the
broader modeling and inference space of Table~\ref{tab:comparison}, in
particular Markov models, flexible random-effects distributions, native
neural-network components, and exploratory development under a unified
likelihood-and-Bayesian interface. For fully bespoke Bayesian models
outside the mixed-effects setting, users can access the underlying
\pkg{Turing.jl} backend directly \citep{turing, turingjl}.

\section{Conclusions}\label{sec:conclusions}
We introduced \pkg{NoLimits.jl}, an open-source \proglang{Julia} framework that
places classical mixed-effects models, mechanistic systems, 
and hybrid mechanistic--neural formulations under a single composable modeling and
inference interface. A macro-based language specifies observation models as
parameterized conditional distributions whose random effects are drawn from the
\pkg{Distributions.jl} ecosystem and may be Gaussian, heavy-tailed, or flow-based
with covariate-dependent parameters, while neural networks, differentiable soft
trees, and differentiable \gls{ode} solvers act as first-class building blocks for
any part of a model. Laplace approximation, the stochastic \gls{em} algorithms
\gls{saem} and \gls{mcem}, maximum likelihood, and full Bayesian inference through
\pkg{Turing.jl} are all reachable from the same specification, together with
simulation, uncertainty quantification, predictive checks, diagnostics, and
cross-validation. We demonstrated this breadth on a warfarin population
pharmacokinetic model, a learnable concentration-effect function for the
pharmacodynamics of the same drug, and normalizing-flow random effects in a
fish-growth study. Across the three studies, the flexible components proved
valuable less for improving individual-level prediction than for revealing
structure that rigid parametric assumptions obscure, namely the learned
concentration-effect shape in the pharmacodynamic example and the recovered
bimodal random-effects distribution in the growth model.

To the best of our knowledge, \nl{} is the only open-source \gls{nlme} framework that
combines \gls{ode} and Markov model classes, heavy-tailed and flow-based
random effects, native neural-network components with automatic differentiation, and a unified
likelihood-and-Bayesian inference interface, broadening the modeling and inference
design space relative to the established packages \pkg{nlmixr2} and \pkg{saemix}.
This breadth comes at the cost of maturity. \nl{} does not yet match their extensive
validation record or large pharmacometrics ecosystems, and it currently lacks some
of their workflow tooling, including bootstrap-based parameter uncertainty,
automated stepwise covariate selection, and import and export of \pkg{NONMEM} and
\pkg{Monolix} models.

Closing these gaps, broadening the catalog of built-in model components, and growing
the body of validation studies are the focus of ongoing development. The default
forward-mode automatic differentiation, required by the nested optimization within the
Laplace and \gls{focei} estimators, remains efficient up to roughly one hundred
parameters and so covers many practical applications, while substantially larger models
would benefit from reverse-mode differentiation, for which planned variational inference
over the random effects \citep{arruda2024amortized, tarek2026fitting} is a promising route.
Further plans include privacy-preserving analysis such as federated learning
\citep{mcmahan2017communication} via DataSHIELD \citep{wilson2017datashield}. \nl{} is released under the MIT license and
developed openly at \url{https://github.com/manuhuth/NoLimits.jl}. We invite the community to apply the
framework to new problems, report issues, and contribute model components, inference
procedures, and examples.

\section*{Acknowledgments}

This study received financial support by the German Research Foundation (Deutsche
Forschungsgemeinschaft, DFG) under Germany's Excellence Strategy (EXC 2047 - 390685813 and
EXC 2151 - 390873048), the University of Bonn (via the Schlegel Professorship of JH), and by
the European Union. Views and opinions expressed are however those of the author(s) only and
do not necessarily reflect those of the European Union or the European Research Council
Executive Agency. Neither the European Union nor the granting authority can be held
responsible for them. This work is supported by ERC grant INTEGRATE, grant agreement number
101126146.

The funders had no role in the study design, data collection, data analyses, data
interpretation, writing, or submission of this manuscript.

\section*{Author contributions}

M.H.\ conceptualized the framework, designed and implemented the \nl{} software package, and
wrote the manuscript. V.W.\ implemented the directed-acyclic-graph-based improvements for the
Markov models. R.G.\ assisted with the theory for the hidden Markov model implementation and
tested the hidden Markov model examples. N.S.\ conducted universal differential equation
experiments and created Figure~\ref{fig:overview}. C.P.\ contributed real-world ordinary
differential equation benchmarks during development. J.A.\ provided methodological input on
normalizing flows, tested and benchmarked the corresponding capabilities. J.H.\ acquired
funding and supervised the work. All authors read and approved the final manuscript.

\section*{Code availability and reproducibility}

\nl{} is open-source software released under the MIT license and developed
publicly at \url{https://github.com/manuhuth/NoLimits.jl}. The results reported
in this article were produced with \nl{} version 0.1.0 under \proglang{Julia}
1.12.1.

All material required to reproduce the results is archived on Zenodo at
\url{https://doi.org/10.5281/zenodo.20797370}. The archive contains the
replication script \code{replication.jl} together with the \code{Project.toml}
and \code{Manifest.toml} files that pin the exact versions of all dependencies.
Executing the script in the provided environment regenerates every figure
reported in this article.

\section*{Declaration of generative AI}
\nl{} was developed with substantial assistance from large language models, primarily Anthropic's Claude (via Claude Code), for code generation, refactoring, test authoring, and documentation. All contributions were reviewed, tested, and are understood by the maintainers, who take full responsibility for the package. This is disclosed per the Julia General Registry's guidance on AI-assisted packages.

During the preparation of this manuscript, the authors used Claude and ChatGPT
to correct grammar and spelling and to improve the style of writing. After using
these tools, the authors reviewed and edited the content as needed and take full
responsibility for the content of this manuscript.

\bibliography{bibfile}

\clearpage
\setcounter{table}{0}
\renewcommand{\tablename}{Supplementary Table}
\renewcommand{\thetable}{S\arabic{table}}
\setcounter{figure}{0}
\renewcommand{\figurename}{Supplementary Figure}
\renewcommand{\thefigure}{S\arabic{figure}}

\begin{appendix}

\section{Further modeling features}\label{app:features}

This appendix develops five features that the main text summarized:
random effects at several grouping levels, multiple observations
per individual, censored observations, Markov observation models,
and the recording of doses and events. The first four extend the
conditional model or the random-effect structure of Section~\ref{sec:nlme} while leaving the estimation
machinery of Section~\ref{sec:point_est} unchanged, and the last describes how
interventions are encoded in the data table that binds a model to its data.
\subsection{Nested random effects}\label{app:nested}

For clarity, the main text considers a single random-effect grouping variable, e.g., the individual level. Many applications are hierarchical, however, with
variability arising at multiple grouping levels, such as individuals
nested within sites or studies. \pkg{NoLimits.jl} supports such
structures by associating each random effect with its own grouping
factor.

Consider individuals $i \in \mathcal{I}_s$ nested within sites
$s = 1, \ldots, S$. A site-level random effect $U_s$ is shared by all
individuals at a site, while an individual-level random effect $B_i$
varies within sites,
\begin{equation}
  U_s \sim p_U(\cdot \mid \theta),
\end{equation}
\begin{equation}
  B_i \sim p_B(\cdot \mid \theta),
  \qquad i \in \mathcal{I}_s,
\end{equation}
\begin{equation}
  Y_{i,j} \mid U_s = u_s, B_i = b_i
  \sim
  p_{Y \mid U,B}(\cdot \mid u_s, b_i, x_i, \theta).
\end{equation}

The marginal likelihood then factorizes over sites rather than
individuals,
\begin{equation}
  p_Y(y \mid \theta)
  =
  \prod_{s=1}^{S}
  \int
  \Biggl[
    \prod_{i \in \mathcal{I}_s}
    \int
    \prod_j
    p_{Y \mid U,B}(y_{i,j} \mid u, b, x_i, \theta)
    p_B(b \mid \theta)\, \mathrm{d}b
  \Biggr]
  p_U(u \mid \theta)\, \mathrm{d}u.
\end{equation}

\pkg{NoLimits.jl} supports both nested and crossed grouping factors
\citep{pinheiro2006efficient}. The estimators of
Section~\ref{sec:point_est} extend naturally by operating on the joint
random-effect vector associated with each grouping block. For Laplace
methods, conditional modes and Hessians are computed jointly over all
random effects in a block. For \gls{em}-type methods, expectations are
taken with respect to the corresponding joint posterior. Full
\gls{mcmc} inference requires only augmenting the hierarchy with the
additional random-effect levels and sampling the resulting joint
posterior. For \gls{ode}-based models, random effects that enter the
dynamics must currently be defined at the individual level.

\subsection{Multiple measurements}\label{app:multi}

Individuals may contribute multiple outcomes rather than a single
response. Let $y_{i,j}^{(r)}$ denote observation $j$ of outcome $r$ for
individual $i$. Conditional on the random effect $B_i$, the likelihood
factorizes as
\begin{equation}
  p_{Y \mid B}(y_i \mid b_i, x_i, \theta)
  =
  \prod_{r=1}^{R}
  \prod_j
  p_{Y \mid B}^{(r)}
  \bigl(
    y_{i,j}^{(r)}
    \mid
    b_i, x_i, \theta
  \bigr),
\end{equation}
where each outcome may follow a different observation model. Sharing the
random effect across outcomes induces dependence between them and enables
joint modeling. Since only the conditional likelihood changes, all
estimators of Section~\ref{sec:point_est} apply without modification. In
\pkg{NoLimits.jl}, multiple outcomes are specified through separate
observation formulas, with missing measurements represented as
\code{missing}.

\subsection{Censored measurements}\label{app:censor}

Censored measurements arise when the exact observation is unknown but is
known to lie within an interval, for example below a lower limit of
quantification. Let $[\ell_{i,j}, u_{i,j}]$ denote the censoring interval
for observation $j$ of individual $i$. Instead of contributing a density,
a censored observation contributes the corresponding interval
probability,
\begin{equation}
  P\bigl(
    \ell_{i,j} \le Y_{i,j} \le u_{i,j}
    \mid B_i = b_i, x_i, \theta
  \bigr)
  =
  P_{Y \mid B}(u_{i,j} \mid b_i, x_i, \theta)
  -
  P_{Y \mid B}(\ell_{i,j} \mid b_i, x_i, \theta),
\end{equation}
where $P_{Y \mid B}$ denotes the conditional cumulative distribution
function. The conditional likelihood is obtained by replacing censored
observations with these interval probabilities, and the resulting
likelihood is marginalized over the random effects as in
Equation~\ref{eq:marginal}.

Likelihood-based estimators evaluate this likelihood directly, whereas
stochastic estimators such as \gls{mcem} and \gls{saem} treat censored
observations as latent variables and sample them from the corresponding
truncated observation model \citep{wei1990monte}. In \pkg{NoLimits.jl},
censoring is specified by wrapping the observation distribution and
providing censoring limits as constants or covariates.

\subsection{Markov observation models}\label{app:hmm}

\begin{table}[t!]
  \centering
  \begin{tabularx}{\linewidth}{@{}lllX@{}}
    \toprule
    Outcome distribution & Time & States & Observation \\
    \midrule
    \code{DiscreteTimeDiscreteStatesHMM}
      & Discrete
      & Hidden
      & Univariate emission \\
    \code{ContinuousTimeDiscreteStatesHMM}
      & Continuous
      & Hidden
      & Univariate emission \\
    \code{MVDiscreteTimeDiscreteStatesHMM}
      & Discrete
      & Hidden
      & Multivariate emissions with missing entries \\
    \code{MVContinuousTimeDiscreteStatesHMM}
      & Continuous
      & Hidden
      & Multivariate emissions with missing entries \\
    \code{DiscreteTimeObservedStatesMarkovModel}
      & Discrete
      & Observed
      & State label \\
    \code{ContinuousTimeObservedStatesMarkovModel}
      & Continuous
      & Observed
      & State label \\
    \code{CoarsedObservedStatesMarkovModel}
      & Both
      & Observed
      & Set of compatible state labels \\
    \bottomrule
  \end{tabularx}
    \caption{Markov-model outcome distributions available in
\pkg{NoLimits.jl}, including hidden- and observed-state models,
discrete- and continuous-time variants, multivariate extensions, and
coarsened-state observations.}
  \label{tab:markov}
\end{table}
\pkg{NoLimits.jl} implements a family of Markov observation models that covers discrete and
continuous time, hidden and directly observed states, univariate and multivariate emissions, and
coarsened state observations (Table~\ref{tab:markov}). For observed-state models the conditional
density $p_{Y \mid B}(y_i \mid b_i, x_i, \theta)$ of Equation~\ref{eq:nlme_obs} is a product of
transition probabilities and is available in closed form. For an \gls{hmm} it is instead obtained by
a recursion over an unobserved within-subject state sequence \citep{rabiner1989tutorial, zucchini2016hidden}.
Let the latent states $Z_{i,1}, \ldots, Z_{i,n_i}$ be random variables taking values in $\{1, \ldots, K\}$
with initial distribution $\pi$, transition matrices $\mathbf{P}_{i,j}$ that may depend on the covariates,
the time, and the random effect, and state-conditional emission densities
$p_{Y \mid Z}(y_{i,j} \mid z_{i,j}, x_{i,j}, \theta)$.
The conditional likelihood is computed by the forward recursion
\begin{align}
  \alpha_{i,1}(k) &= \pi_k\, p_{Y \mid Z}(y_{i,1} \mid k, x_{i,1}, \theta), \nonumber\\
  \alpha_{i,j}(l) &= \Bigl[ \sum_{k=1}^{K} \alpha_{i,j-1}(k)\, [\mathbf{P}_{i,j}]_{kl} \Bigr]\,
    p_{Y \mid Z}(y_{i,j} \mid l, x_{i,j}, \theta), \nonumber\\
  p_{Y \mid B}(y_i \mid b_i, x_i, \theta) &= \sum_{k=1}^{K} \alpha_{i,n_i}(k),
\end{align}
which enters the joint and marginal densities of Equations~\ref{eq:joint} and \ref{eq:marginal} in place of a
closed-form term, so that every estimator of Section~\ref{sec:point_est} applies without change.
In the continuous-time variants the transition matrices are obtained as matrix exponentials of a rate matrix.
The transition matrix may be parameterized by a neural network mapped through a row-wise softmax and the most likely hidden-state sequence is recovered by
Viterbi decoding.

For example, an \gls{hmm} with two hidden states is defined as follows.
\begin{CodeInput}
model = @Model begin
    @fixedEffects begin
        p12   = RealNumber(0.1, scale = :logit)
        p21   = RealNumber(0.2, scale = :logit)
        mu1   = RealNumber(0.0)
        mu2   = RealNumber(3.0)
        sigma = RealNumber(0.5, scale = :log)
    end
    @formulas begin
        P = [1 - p12      p12;
             p21      1 - p21]
        y ~ DiscreteTimeDiscreteStatesHMM(P,
            (Normal(mu1, sigma), Normal(mu2, sigma)),
            Categorical([0.5, 0.5]))
    end
end
\end{CodeInput}
The three arguments of \code{DiscreteTimeDiscreteStatesHMM} specify the
transition matrix, the state-specific emission distributions, and the
initial state distribution. While this explicit construction is
convenient for small state spaces, larger models can declare the full
transition matrix and initial distribution directly using the
\code{DiscreteTransitionMatrix} and \code{ProbabilityVector}
parameters of Table~\ref{tab:paramtypes}. For continuous-time models, the
transition structure is instead supplied as the rate matrix
\code{ContinuousTransitionMatrix}.

\subsection{Laplace and FOCEI approximation}\label{app:laplace}

Let
\begin{equation}
  g_i(b)
  =
  \log p_{Y \mid B}(y_i \mid b, x_i, \theta)
  +
  \log p_B(b \mid x_i, \theta),
\end{equation}
so that the marginal likelihood contribution of individual $i$ is
$\int_{\mathcal{B}} \exp(g_i(b))\, \mathrm{d}b$. A second-order Taylor
expansion around the \gls{ebe} $\hat{b}_i$ yields
\begin{equation}
  g_i(b)
  \approx
  g_i(\hat{b}_i)
  +
  \tfrac{1}{2}
  (b - \hat{b}_i)^\top
  H_i(\theta)
  (b - \hat{b}_i),
  \qquad
  H_i(\theta)
  =
  \nabla_b^2 g_i(b)\big|_{b=\hat{b}_i},
  \label{eq:hessian}
\end{equation}
where the first-order term vanishes because $\hat{b}_i$ is a mode.
Substituting this approximation into the marginal integral yields the
Laplace approximation of Equation~\ref{eq:laplace}
\citep{tierney1986accurate}.

The \gls{focei} approximation uses the same form but replaces the exact
Hessian by the curvature implied by a first-order linearization of the
conditional model around $\hat{b}_i$
\citep{wang2007derivation, pinheiro2006efficient}. Letting $\phi_{ij}(b)$
denote the parameters of the observation distribution for observation $j$ of
individual $i$, \gls{focei} approximates the curvature of the conditional
log-likelihood by its Fisher information \citep{fisher1925theory} while
retaining the exact curvature of the random-effects distribution,
\begin{equation}
  H_i(\theta)
  =
  -\sum_{j=1}^{n_i}
  J_{ij}^\top \mathcal{I}(\phi_{ij}) J_{ij}
  +
  \nabla_b^2
  \log p_B(\hat{b}_i \mid x_i, \theta),
  \qquad
  J_{ij}
  =
  \frac{\partial \phi_{ij}}{\partial b}
  \bigg|_{b=\hat{b}_i},
  \label{eq:focei}
\end{equation}
where $\mathcal{I}(\phi_{ij})$ is the Fisher information of the observation
distribution and $J_{ij}$ is the Jacobian of its parameters with respect to
the random effects. For bounded
random-effect domains, the Gaussian integral is evaluated over the
corresponding support rather than over $\mathbb{R}^{n_b}$
\citep{greenberg2013foundations}.

\paragraph{Gradient via the envelope theorem.}
Gradient-based optimization of $\theta$ requires the gradient of the approximate marginal log-likelihood.
Making the dependence on $\theta$ explicit and taking the logarithm of Equation~\ref{eq:laplace}, the per-individual approximate marginal log-likelihood is
\begin{equation}
  \ell_i(\theta)
  = g_i\bigl(\hat{b}_i(\theta), \theta\bigr)
    - \tfrac{1}{2} \log\det\bigl(-H_i(\theta)\bigr)
    + \tfrac{n_b}{2}\log(2\pi),
  \label{eq:laplace_ll}
\end{equation}
where $\hat{b}_i(\theta) = \argmax_b g_i(b, \theta)$ is the \gls{ebe} and
$H_i(\theta) = \nabla_b^2 g_i(b, \theta)|_{b = \hat{b}_i}$. Both $\hat{b}_i(\theta)$ and $H_i(\theta)$ depend
on $\theta$, so a naive total derivative would have to differentiate through the inner optimization. The
envelope theorem removes this for the first term, because $\hat{b}_i(\theta)$ maximizes $g_i(\cdot, \theta)$
the inner gradient $\nabla_b g_i(\hat{b}_i, \theta)$ vanishes, and the total derivative of
$g_i(\hat{b}_i(\theta), \theta)$ reduces to its explicit partial $\nabla_\theta g_i(\hat{b}_i, \theta)$. The
log-determinant term is not stationary in $b$, so its total derivative retains the sensitivity of the mode,
\begin{equation}
  \nabla_\theta \ell_i(\theta)
  = \nabla_\theta g_i(\hat{b}_i, \theta)
    - \tfrac{1}{2} \Bigl[
        \nabla_\theta \log\det(-H_i)
        + \bigl( \nabla_b \log\det(-H_i) \bigr)^\top
          \frac{\mathrm{d}\hat{b}_i}{\mathrm{d}\theta}
      \Bigr],
  \label{eq:laplace_grad}
\end{equation}
where the partial derivatives of the log-determinant treat $b$ and $\theta$ as independent. The mode
sensitivity follows from the implicit function theorem applied to the optimality condition
$\nabla_b g_i(\hat{b}_i(\theta), \theta) = 0$,
\begin{equation}
  \frac{\mathrm{d}\hat{b}_i}{\mathrm{d}\theta}
  = \bigl(-H_i(\theta)\bigr)^{-1}\, \nabla_\theta \nabla_b g_i(\hat{b}_i, \theta).
  \label{eq:mode_sensitivity}
\end{equation}
\pkg{NoLimits.jl} evaluates Equations~\ref{eq:laplace_grad} and \ref{eq:mode_sensitivity} by automatic
differentiation of $g_i$ and $\log\det(-H_i)$ at the converged mode, reusing the Cholesky factor of
$-H_i(\theta)$ already formed for Equation~\ref{eq:laplace}. The same identities apply to \gls{focei} with the respective approximation of the hessian term (Equation~\ref{eq:focei}).
\subsection{EM algorithm and Monte Carlo expectation}\label{app:em}

The \gls{em} algorithm constructs the surrogate objective
$Q(\theta \mid \theta_t)$ of Equation~\ref{eq:qfun}, whose maximization
guarantees non-decreasing marginal likelihood values
\citep{dempster1977maximum}. For nonlinear and non-Gaussian models, the
required expectation is generally intractable. \gls{mcem} therefore
approximates it using samples
$b_i^{(1)}, \ldots, b_i^{(M_t)}$ from the posterior distribution of
Equation~\ref{eq:re_post},
\begin{equation}
  Q(\theta \mid \theta_t)
  \approx
  \sum_{i=1}^{N}
  \frac{1}{M_t}
  \sum_{m=1}^{M_t}
  \log p_{Y,B}
  \bigl(
    y_i, b_i^{(m)}
    \mid
    x_i, \theta
  \bigr).
\end{equation}

In \pkg{NoLimits.jl}, these samples are obtained through
\pkg{Turing.jl}, typically using Metropolis--Hastings or NUTS sampling
to target the posterior distribution of
Equation~\ref{eq:re_post}
\citep{robert2004monte, chib1995understanding}.
\subsection{SAEM learning-rate schedule}\label{app:gamma}

The \gls{saem} learning rate $\gamma_k$ controls the update of the
sufficient statistics of Equation~\ref{eq:saem} at iteration $k$. \pkg{NoLimits.jl} uses a
piecewise schedule consisting of an optional burn-in, an initial
plateau, and two polynomial decay phases,
\begin{equation}
\gamma_k =
\begin{cases}
0,
& 1 \le k \le K_{\mathrm{burn}},
\\[4pt]
1,
& K_{\mathrm{burn}} < k \le K_{\mathrm{burn}} + K_0,
\\[4pt]
k^{-a_1},
& K_{\mathrm{burn}} + K_0 < k \le K_{\mathrm{burn}} + K_0 + K_1,
\\[4pt]
\bigl(
k - K_{\mathrm{offset}}
+ \gamma_{K_{\mathrm{offset}}}^{-1/a_2}
\bigr)^{-a_2},
& k > K_{\mathrm{burn}} + K_0 + K_1,
\end{cases}
\label{eq:saem_gamma}
\end{equation}
where
\begin{equation}
K_{\mathrm{offset}}
=
K_{\mathrm{burn}} + K_0 + K_1.
\end{equation}

The first phase performs no updates, the second uses full updates, and
the final two phases gradually reduce the learning rate. The last branch
ensures a continuous transition between the two decay regimes while
preserving polynomial decay. For $a_1, a_2 \in (0.5, 1]$, the schedule
satisfies the Robbins--Monro conditions required for stochastic
approximation convergence.

\subsection{SAEM sufficient-statistic update}\label{app:saem}

Suppose the complete-data density belongs to an exponential family,
\begin{equation}
  \log p_{Y,B}(y_i,b \mid x_i,\theta)
  =
  \langle \phi(\theta), T(y_i,b) \rangle
  - \psi(\theta)
  + \mathrm{const},
\end{equation}
where $T$ is a sufficient statistic, $\phi(\theta)$ the natural
parameter, and $\psi(\theta)$ the log-partition function. The surrogate
of Equation~\ref{eq:qfun} then depends on $\theta$ only through the
expected sufficient statistics. SAEM replaces these expectations by the
stochastic recursion
\begin{equation}
  \bar{T}^{(t)}
  =
  (1 - \gamma_t)\,\bar{T}^{(t-1)}
  +
  \gamma_t
  \sum_{i=1}^{N}
  \frac{1}{M_t}
  \sum_{s=1}^{M_t}
  T\bigl(y_i, b_i^{(t,s)}\bigr),
  \label{eq:saem_suff}
\end{equation}
where $b_i^{(t,s)}$ are samples from the conditional distribution of
the random effects. The maximization step then updates the parameters by
maximizing
\begin{equation}
  \langle \phi(\theta), \bar{T}^{(t)} \rangle
  -
  N \psi(\theta),
\end{equation}
which is available in closed form for common exponential-family models
such as Gaussian random effects
\citep{kuhn2005maximum, comets2017parameter}.

In \pkg{NoLimits.jl}, the random-effect samples are generated by the
default \code{SaemixMH} sampler, a lightweight Metropolis--Hastings
scheme inspired by \pkg{saemix} \citep{comets2017parameter} that avoids
constructing a full \pkg{Turing.jl} model at each iteration.

\subsection{Prediction}\label{app:prediction}

This appendix gives the predictors summarized in
Section~\ref{sec:prediction}, which depend on whether the individual was
observed during model fitting.

\paragraph{Individuals observed in training.}
For an individual included in the training data, the observed responses
$y_i$ inform the posterior of the random effect in
Equation~\ref{eq:re_post}. A computationally inexpensive prediction at
covariate value $x_i$, possibly a new value or time point, substitutes the
corresponding \gls{ebe} $\hat{b}_i$ into the conditional mean,
\begin{equation}
  \hat{y}_i
  =
  \Ex\bigl(
    Y_i
    \bigm|
    B_i = \hat{b}_i,
    x_i,
    \theta
  \bigr),
  \label{eq:pred_ebe}
\end{equation}
whereas propagating the uncertainty in the random effect averages the
conditional mean over the posterior,
\begin{equation}
  \hat{y}_i
  =
  \int_{\mathcal{B}}
  \Ex\bigl(
    Y_i
    \mid
    B_i = b,
    x_i,
    \theta
  \bigr)\,
  p_{B \mid Y}(b \mid y_i, x_i, \theta)
  \,\mathrm{d}b,
  \label{eq:pred_posterior}
\end{equation}
which reduces to the plug-in predictor of Equation~\ref{eq:pred_ebe} when the
conditional mean is linear in the random effect
\citep{lindstrom1990nonlinear}.

\paragraph{Individuals not observed in training.}
For an unseen individual, no posterior distribution is available and two
prediction strategies arise. The first is the plug-in predictor, which
replaces the random effect by its covariate-dependent prior mean,
\begin{equation}
  \hat{y}_i
  =
  \Ex\bigl(
    Y_i
    \bigm|
    B_i = \Ex(B_i \mid x_i, \theta),
    x_i,
    \theta
  \bigr),
  \label{eq:pred_plugin}
\end{equation}
while the second is the marginal predictor,
\begin{equation}
  \hat{y}_i
  =
  \int_{\mathcal{B}}
  \Ex\bigl(
    Y_i
    \mid
    B_i = b,
    x_i,
    \theta
  \bigr)
  p_B(b \mid x_i, \theta)
  \,\mathrm{d}b,
  \label{eq:pred_marginal}
\end{equation}
which corresponds to the population mean implied by the marginal model in
Equation~\ref{eq:marginal} \citep{lavielle2014mixed}. The two predictors
coincide when the conditional mean is linear in the random effects and
generally differ otherwise.

\subsection{Numerical configuration}\label{app:config}

Beyond the optimizer and parameter transformations of
Section~\ref{sec:est_nl}, two further numerical components are configured
separately from the model and its data binding, the multistart procedure
and the \gls{ode} solver.

\paragraph{Multistart.}
To improve robustness against local optima, any optimization-based estimator
can be wrapped in the \code{Multistart} procedure, which draws starting points
from user-specified distributions either independently or through \gls{lhs}
\citep{mckay1979comparison}. Candidate starts are first ranked by a cheap
screening objective and only the most promising are fully optimized,
controlled by \code{n\_draws\_requested} and \code{n\_draws\_used}, and the
wrapper is combined with an estimator by passing it as the first argument to
\code{fit\_model}.

\begin{CodeInput}
ms = Multistart(dists = (beta1 = Normal(0.0, 1.0),
                         beta2 = Normal(1.0, 0.5)),
                n_draws_requested = 200, n_draws_used = 50,
                sampling = :lhs)
res  = fit_model(ms, dm, Laplace())
best_fit = get_multistart_best(res)
plot_multistart_waterfall(res)
\end{CodeInput}

Fixed effects without a sampling distribution are held at their initial
values, the best fit is recovered with \code{get\_multistart\_best}, and the
spread of objective values across starts is summarized by the multistart
waterfall plot.

\paragraph{ODE solver configuration.}
ODE solver settings are configured separately from the model
specification using the function \code{set\_solver\_config}. This keeps the numerical
integrator independent of both the data binding and the estimation
method. By default, \nl{} uses \code{Tsit5()}, but any solver from
\pkg{OrdinaryDiffEq.jl} can be supplied, including stiff solvers,
together with solver-specific options such as \code{abstol} (and
\code{reltol}), which are forwarded to the \gls{ode} solver. For example,
\begin{CodeInput}
model = set_solver_config(model; alg = Tsit5(), kwargs = (; abstol = 1e-8))
\end{CodeInput}
configures the model to use the explicit \code{Tsit5()} solver with a
tighter absolute tolerance. As with all configuration functions in
\nl{}, the call returns an updated \code{Model} object rather than
modifying its argument in place.

\section{Additional model specifications and diagnostics}

\subsection{Warfarin diagnostics}
\label{app:warfarin_extra}

This appendix reports convergence and inter-individual-variability
diagnostics for the warfarin analysis of Section~\ref{sec:warfarin}, together
with the marginal posteriors of the variance components from its Bayesian fit.

\subsubsection{Convergence}

Convergence of the \gls{mcem} fit is assessed from the per-iteration
trajectories of the parameters and surrogate objective $Q$, obtained with
\code{plot\_em\_trajectories}:

\begin{CodeInput}
plot_em_trajectories(fit; ncols = 3, style = my_style)
\end{CodeInput}

\begin{figure}[t!]
  \centering
  \includegraphics[width = \textwidth]{%
    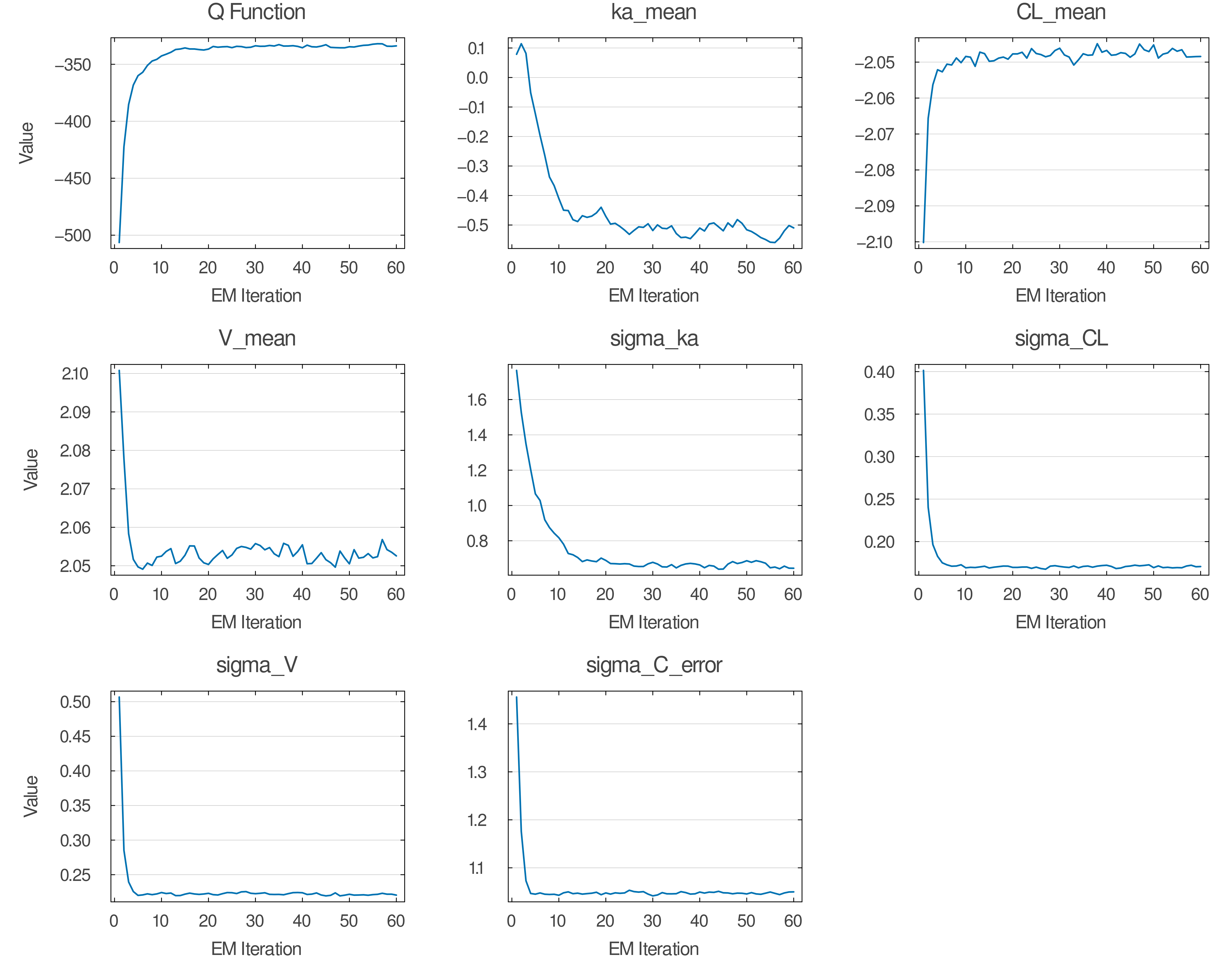}
  \caption{\gls{mcem} convergence trajectories for the warfarin model:
    the Q-function (upper left) and all fixed-effect parameters.}
  \label{fig:warfarin_em}
\end{figure}

Figure~\ref{fig:warfarin_em} shows stable parameter estimates and a
plateauing $Q$-function after roughly 20 iterations, indicating
convergence. The absorption-rate parameter \code{ka\_mean} exhibits the
largest Monte Carlo variability, consistent with its wider confidence
interval.

\subsubsection{Inter-individual variability}
\begin{figure}[t!]
  \centering
  \includegraphics[width = \textwidth]{%
    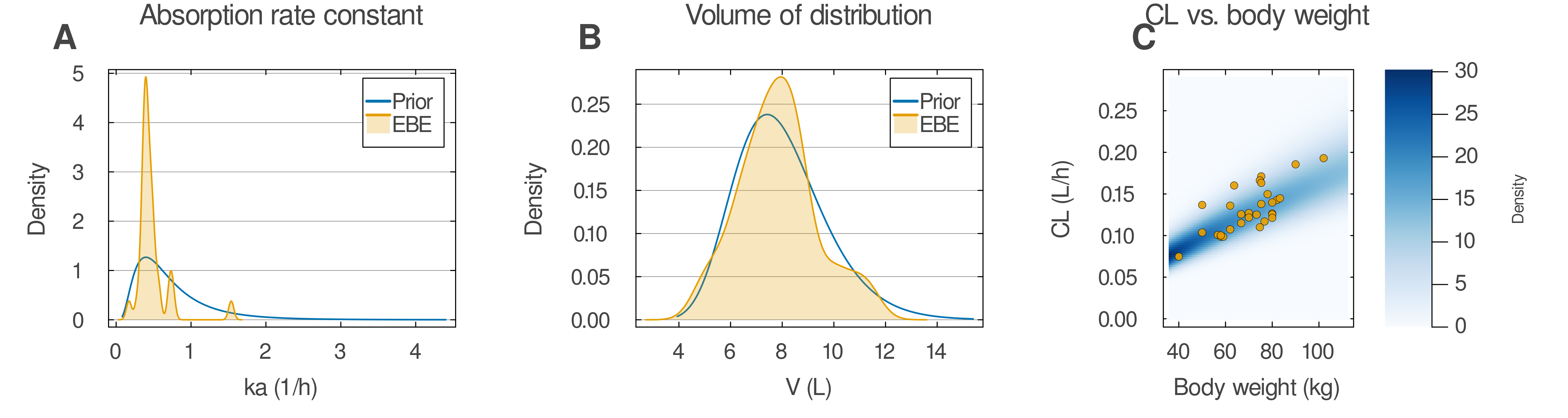}
  \caption{Inter-individual variability diagnostics. (A) Population
    prior and \gls{kde} of \glspl{ebe} for \code{ka}. (B) Same for
    \code{V}. (C) Conditional distribution of \code{CL} as a function
    of body weight with individual \glspl{ebe} overlaid.}
  \label{fig:warfarin_iiv}
\end{figure}
Figure~\ref{fig:warfarin_iiv} compares the fitted random-effects
distributions with the corresponding \glspl{ebe}. For \code{V}, the
individual estimates closely follow the population distribution
(Figure~\ref{fig:warfarin_iiv}B), whereas the \code{ka} estimates are
more concentrated around the population mean
(Figure~\ref{fig:warfarin_iiv}A), indicating shrinkage.

Shrinkage, defined in Section~\ref{sec:evaluation}, is quantified with \code{compute\_shrinkage}:
\begin{CodeInput}
shrink = compute_shrinkage(fit)
\end{CodeInput}
yielding
\begin{CodeOutput}
ETA shrinkage
  ka : 42.9%
  CL : 15.6%
  V  : 11.7%
\end{CodeOutput}

The absorption-rate parameter \code{ka} exhibits substantial shrinkage
(42.9\%), exceeding the conventional 30\% threshold
\citep{savic2009importance} and reflecting the limited information in
the absorption phase. In contrast, \code{CL} and \code{V} show low
shrinkage (15.6\% and 11.7\%), indicating that the elimination phase
provides sufficient individual information. Figure~\ref{fig:warfarin_iiv}C
also confirms the expected increase of clearance with body weight under
the allometric model.

\subsubsection{Posterior distributions of the standard deviations}
\label{app:warfarin_bayes}

\begin{figure}[t!]
  \centering
  \includegraphics[width = \textwidth]{%
    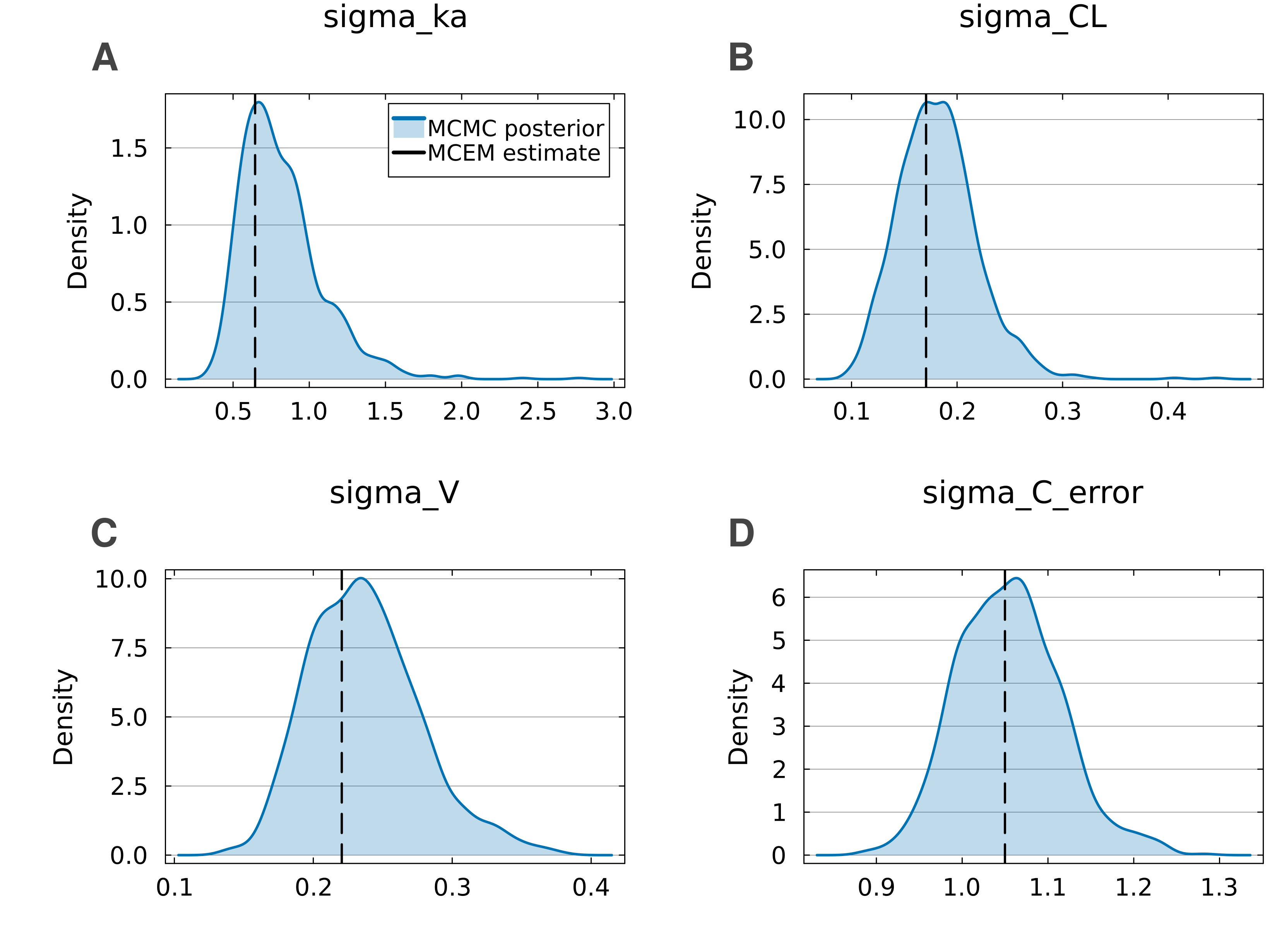}
  \caption{Marginal posterior distributions of the random-effect standard
    deviations and the residual standard deviation from the Bayesian fit of the
    warfarin model, with the corresponding \gls{mcem} point estimates as dashed
    lines: (A)~$\sigma_{k_a}$, (B)~$\sigma_{\mathrm{CL}}$, (C)~$\sigma_V$,
    (D)~$\sigma_C$.}
  \label{fig:warfarin_mcmc_sigmas}
\end{figure}

Figure~\ref{fig:warfarin_mcmc_sigmas} compares the posteriors of the four
standard-deviation parameters from the Bayesian fit of
Section~\ref{sec:warfarin_bayes} with their \gls{mcem} point estimates. Each
estimate lies close to the mode of its posterior. The standard deviations of
clearance, volume, and the residual error have tight posteriors, whereas the
absorption-rate standard deviation $\sigma_{k_a}$ has a broader, right-skewed
posterior, mirroring the behavior of $\mu_{k_a}$.

\subsection{Parametric pharmacodynamic reference model}
\label{app:warfarin_pkpd_param}

The \gls{pd} example of Section~\ref{sec:warfarin_pkpd} compares two learnable
concentration-effect functions against a parametric saturable \code{Emax}
reference. The neural-network and soft-tree models are listed in full in that
section. For completeness, the \code{@Model} specification of the parametric
reference is given here. It shares the \gls{pk} submodel, random-effects
structure, and observation model of the neural-network model, differing only in
the fixed effects and the effect function. The fixed effects carry the saturable
parameters \code{Emax} and \code{EC50} in place of the learnable
\code{ml\_params}, and the effect enters the turnover model of
Equation~\ref{eq:pd_turnover} through the saturable form of
Equation~\ref{eq:effect} rather than a learned function:

\begin{CodeInput}
model_par = @Model begin
    @fixedEffects begin
        ka_mean    = RealNumber(0.0); CL_mean    = RealNumber(-2.0)
        V_mean     = RealNumber(2.0); kout_mean  = RealNumber(-3.0)
        sigma_ka   = RealNumber(0.3, scale = :log)
        sigma_CL   = RealNumber(0.3, scale = :log)
        sigma_V    = RealNumber(0.3, scale = :log)
        sigma_C    = RealNumber(1.0, scale = :log)
        sigma_kout = RealNumber(0.3, scale = :log)
        sigma_R    = RealNumber(5.0, scale = :log)
        Emax       = RealNumber(1.0, scale = :log)
        EC50       = RealNumber(1.0, scale = :log)
    end
    @covariates begin
        t  = Covariate()
        d  = ConstantCovariate(; constant_on = :id)
        wt = ConstantCovariate(; constant_on = :id)
        R0 = ConstantCovariate(; constant_on = :id)
    end
    @randomEffects begin
        ka   = RandomEffect(LogNormal(ka_mean, sigma_ka); column = :id)
        CL   = RandomEffect(
                   LogNormal(CL_mean + 0.75 * log(wt / 70.0), sigma_CL);
                   column = :id)
        V    = RandomEffect(LogNormal(V_mean, sigma_V); column = :id)
        kout = RandomEffect(LogNormal(kout_mean, sigma_kout); column = :id)
    end
    @DifferentialEquation begin
        eff(t) = Emax * (center / V) / (EC50 + center / V)
        D(depot)    ~ -ka * depot
        D(center)   ~  ka * depot - (CL / V) * center
        D(response) ~  kout * (R0 - response * (1 + eff(t)))
    end
    @initialDE begin
        depot = d; center = 0.0; response = R0
    end
    @formulas begin
        C ~ Normal(center(t) / V, sigma_C); R ~ Normal(response(t), sigma_R)
    end
end
\end{CodeInput}

The cross-validated comparison of Section~\ref{sec:warfarin_pkpd} fits this model
with the same \code{fit\_cv} call used for the learnable variants, so all three
effect models are estimated and scored under an identical protocol.

\end{appendix}

\end{document}